\documentclass[journal]{IEEEtran}

\usepackage{fancyhdr}
\usepackage{epsfig}
\usepackage{threeparttable}
\usepackage{epsf,epsfig}
\usepackage{amsthm}
\usepackage[cmex10]{amsmath}
\usepackage{amssymb, amsfonts}
\usepackage[noadjust]{cite}
\usepackage{dsfont}
\usepackage{subfigure}
\usepackage{color}
\usepackage{soul}
\usepackage{amssymb}
 \usepackage{accents}
  \usepackage{algorithm}
   \usepackage[english]{babel}
    \usepackage{url}
    \usepackage{framed} 

\newtheorem{theorem}{Theorem}

\newtheorem{lemma}{Lemma}

\newtheorem{assumption}{Assumption}

\newtheorem{remark}{\bf Remark}

\def\qed{$\Box$}

\def\proof{\noindent{\emph{Proof:} }}

\def
\endproof{\hspace*{\fill}~\qed
\par
\endtrivlist\unskip}
\def\endproof{\hspace*{\fill}~\qed\par\endtrivlist\vskip3pt}

\def\phi{\varphi}

\def\l{\left}
\def\r{\right}
\def\({\left(}
\def\){\right)}

\def\bg{{\mathbf{g}}}

\def\bv{{\mathbf{v}}}
\def\bw{{\mathbf{w}}}
\def\bx{{\mathbf{x}}}

\def\bz{{\mathbf{z}}}
\def\b0{{\mathbf{0}}}

\def\bL{{\mathbf{L}}}

\newcommand{\var}{\mathsf{var}}

\newcommand{\nn}{\nonumber}

\begin{document}	
\title{\huge One-Bit Over-the-Air Aggregation for Communication-Efficient Federated Edge Learning: Design and Convergence Analysis}
\author{Guangxu Zhu, Yuqing Du, Deniz G\"und\"uz, and  Kaibin Huang  
 \thanks{\noindent G. Zhu is with the Shenzhen Research Institute of Big Data, Shenzhen, China (Email: gxzhu@sribd.cn). Y. Du and K. Huang are with the Dept. of Electrical and Electronic Engineering at The  University of  Hong Kong, Hong Kong (Email: \{yqdu, huangkb\}@eee.hku.hk). Deniz G\"und\"uz is with the Dept. of Electrical and Electronics Engineering at the Imperial College of London, London, UK (Email: d.gunduz@imperial.ac.uk). Corresponding author: K. Huang. Part of this paper has been presented at the IEEE Global Communications Conference (GLOBECOM), Taipei, Taiwan, Dec. 2020.}
 \thanks{\noindent 
The work of G. Zhu was supported in part by National Natural Science Foundation of China under grant 62001310, in part by National Key R\&D Program of China under grant 2018YFB1800800, in part by Guangdong Province Key Area R\&D Program under grant 2018B030338001, and in part by Shenzhen Peacock Plan under grant KQTD2015033114415450.
The work of K. Huang is supported by the Hong Kong Research Grants Council under the Grants 17208319 and 17209917, Innovation and Technology Fund under Grant GHP/016/18GD, and Guangdong Basic and Applied Basic Research Foundation under Grant 2019B1515130003.
The work of  D. G\"und\"uz received funding from the European Research Council (ERC) under Starting Grant BEACON (agreement no. 677854). } 
 }
\maketitle


\begin{abstract}
\emph{Federated edge learning} (FEEL) is a popular framework for model training at an edge server using data distributed at edge devices (e.g., smart-phones and sensors) without compromising their privacy. In the FEEL framework, edge devices periodically transmit high-dimensional stochastic gradients to the edge server, where these gradients are aggregated and used to update a global model. 
When the edge devices share the same communication medium, the multiple access channel (MAC) from the devices to the edge server induces a communication bottleneck. To overcome this bottleneck, an efficient  broadband analog transmission scheme has been recently proposed, featuring the aggregation of analog modulated gradients (or local models) via the waveform-superposition property of the wireless medium. However, the assumed linear analog modulation makes it difficult to deploy this technique in modern wireless systems that exclusively use digital modulation. To address this issue, we propose in this work a novel digital version of broadband over-the-air aggregation, called \emph{one-bit broadband digital aggregation} (OBDA). {The new scheme features one-bit gradient quantization followed by digital \emph{quadrature amplitude modulation} (QAM) at edge devices} and over-the-air majority-voting based decoding at edge server. We provide a comprehensive analysis  of the effects of wireless channel hostilities (channel noise, fading, and channel estimation errors) on the convergence rate of the proposed FEEL scheme. The analysis shows that the hostilities slow down the convergence of the learning process by introducing a scaling factor and a bias term into the gradient norm. However, we show that all the negative effects vanish as the number of participating devices grows, but at a different rate for each type of channel hostility. 
\end{abstract}


\section{introduction}
Edge  learning refers to the deployment of learning algorithms at the network edge so as to have rapid access to massive mobile data for training \emph{artificial intelligence} (AI) models~\cite{zhu2018towards,mao2017survey,wang2018edge}. 
A popular framework, called \emph{federated edge learning} (FEEL), distributes the task of model training over many edge devices \cite{mcmahan2017communication,konevcny2016federated}. Thereby, FEEL exploits distributed data without compromising their privacy, and furthermore leverages the computation resources of edge devices. Essentially, the framework is a distributed implementation of \emph{stochastic gradient descent} (SGD) in a wireless network. 
In FEEL, edge devices periodically upload locally trained models to an edge server, which are then aggregated and used to update a global model. When the edge devices participating in the learning process share the same wireless medium to convey local updates to the edge server, the limited radio resources can cause severe congestion over the air interface, resulting in a communication bottleneck for FEEL. One promising solution is \emph{over-the-air aggregation} also called \emph{over-the-air computation} (AirComp) that exploits the waveform superposition property of the wireless medium to support simultaneous transmission by all the devices~\cite{zhu2018low,amiri2019machine}. Compared with  conventional orthogonal access schemes, this can dramatically reduce the required resources, particularly many devices participate in FEEL. However, the uncoded linear analog modulation scheme used for over-the-air aggregation in \cite{zhu2018low,amiri2019machine} may cause difficulty in their deployment in existing systems, which typically use digital modulation (e.g., 3GPP).
In this work, we propose a new design, called \emph{one-bit broadband digital aggregation} (OBDA), that extends the conventional analog design by featuring digital modulation of gradients to facilitate efficient implementation of FEEL in a practical wireless system. We present a comprehensive convergence analysis of OBDA to quantify the effects of channel hostilities, including channel noise, fading and estimation errors, on its convergence rate. The results provide guidelines on coping with the channel hostilities in FEEL systems with OBDA.

\subsection{FEEL over a Multiple Access Channel (MAC)}
A typical FEEL algorithm involves iterations between two steps as shown in Fig. \ref{Fig:1}. 
In the first step, the edge server broadcasts the current version of the global model to the participating edge devices. Each edge device employs SGD using only locally available data. In the next step, edge devices convey their local updates (gradient estimates or model updates) to the edge server.
Each iteration of these two steps is called one \emph{communication round}. The iteration continues until the global model converges. 

 The communication bottleneck has already been acknowledged as a major challenge
  in the federated learning literature, and several strategies have been proposed to reduce the communication overhead. We can identify  three main approaches. The first is to discard the updates from slow-responding edge devices (stragglers) for fast update synchronization~\cite{chen2016revisiting,tandon2017gradient}. Another approach is to employ  update significance rather than the computation speed  to schedule devices~\cite{kamp2018efficient,chen2018lag}. Update significance is measured by either the model variance \cite{kamp2018efficient}, or the gradient divergence~\cite{chen2018lag} corresponding to  model-averaging~\cite{mcmahan2017communication} and gradient-averaging~\cite{konevcny2016federated}  implementation methods, respectively. 
{ The last approach focuses on update compression by exploiting the sparsity of gradient updates~\cite{lin2017deep,sattler2019sparse} and low-resolution gradient/model-parameter quantization \cite{alistarh2017qsgd,2018signsgd,reisizadeh2019fedpaq}. 
 However, all these approaches assume reliable links between the devices and the server and ignore the wireless nature of the communication medium. }

The envisioned implementation of FEEL in practical wireless networks requires taking into account  wireless channel hostilities and the scarcity of radio resources. The first works in the literature that studied FEEL taking into account the physical layer resource constraints focus on over-the-air aggregation \cite{zhu2018low,amiri2019machine,amiri2019federated,amiri2019collaborative,yang2018federated}.
Specifically, a broadband over-the-air aggregation system based on analog modulation called \emph{broadband analog aggregation} (BAA) is designed in \cite{zhu2018low}, where the gradients/models transmitted by devices are averaged over frequency sub-channels. For such a system, several communication-learning tradeoffs are derived to guide the designs of broadband power control and device scheduling. Over-the-air aggregation is also designed for narrow-band channels with an additional feature of gradient dimension reduction exploiting gradient sparsity in \cite{amiri2019machine,amiri2019federated}. Subsequently, \emph{channel-state information} (CSI) requirement for over-the-air aggregation is relaxed by exploiting multiple antennas at the edge server in \cite{amiri2019collaborative}. 
The joint design of device scheduling and beamforming for over-the-air aggregation is investigated in \cite{yang2018federated} to accelerate FEEL in a multi-antenna system. 
Few other recent works focus on radio resource management  \cite{abad2019hierarchical,chen2019joint,yang2019scheduling}. In \cite{abad2019hierarchical}, a hierarchical FEEL framework is introduced in a cellular network. 
 In \cite{chen2019joint}, a novel bandwidth allocation strategy is proposed for minimizing the model training loss of FEEL via convergence analysis accounting for packet transmission errors. Different classic scheduling schemes, such as proportional fair scheduling, are applied in a FEEL system and their effects on the convergence rate are studied in \cite{yang2019scheduling}. 

The performance of FEEL is typically measured in terms of the convergence rate, which quantifies how fast the global model converges over communication rounds. 
The current work  is the first to present a comprehensive framework for convergence analysis targeting FEEL with AirComp.

\subsection{Over-the-Air Aggregation}

With over-the-air aggregation, the edge server receives an approximate  (noisy version) of the desired functional value, efficiently exploiting the available bandwidth by simultaneous transmission from all the devices, as opposed to orthogonalized massive access. The idea of over-the-air aggregation has previously been studied in the context of data aggregation in sensor networks, known also as AirComp. In \cite{nazer2007computation}, function computation over a MAC is studied from a fundamental information theoretic point of view, assuming \emph{independent and identically distributed} (i.i.d.) source sequences, and focusing on the asymptotic computation rate. This is extended in \cite{GoldenbaumTCOM2013} to wireless MACs and a more general class of non-linear functions. A practical implementation is presented in \cite{abari2016over}. 
From an  implementation perspective, techniques for distributed power control and robust design against channel estimation errors are proposed in~\cite{xiao2008linear} and ~\cite{goldenbaum2014channel}, respectively. More recently, AirComp techniques are designed for \emph{multiple-input-multiple-output} (MIMO) systems for enabling spatially multiplexed vector-valued function computation in \cite{zhu2018mimo,li2019wirelessly,wen2019reduced}. To this end, receive beamforming and an enabling scheme for CSI feedback are designed in~\cite{zhu2018mimo}. The framework is extended to wirelessly-powered AirComp~\cite{li2019wirelessly} and massive MIMO AirComp systems \cite{wen2019reduced}.

Existing works on over-the-air aggregation, as well as its application in the context of FEEL \cite{zhu2018low,amiri2019machine,amiri2019federated,amiri2019collaborative,yang2018federated} consider analog modulation assuming that the transmitter can modulate the carrier waveform as desired, freely choosing the I/Q coefficients as arbitrary real number. However, most existing wireless devices come with embedded digital modulation chips, and they may not be capable of employing an arbitrary modulation scheme.
{ In particular, modern cellular systems are based on \emph{orthogonal frequency division multiple access} (OFDMA) using \emph{quadrature amplitude modulation} (QAM). Therefore, the goal of the paper is to extend the over-the-air aggregation framework to FEEL considering transmitters that are limited to QAM.}

\subsection{Contributions and Organization}
In this paper, we consider the implementation of over-the-air aggregation for  FEEL over a practical wireless system with digital modulation. Building on the signSGD proposed in \cite{2018signsgd} and \emph{orthogonal frequency division multiplexing} (OFDM) transceivers, { we design an elaborate FEEL scheme, called OBDA, which features one-bit gradient quantization and QAM modulation at devices, and  over-the-air majority-vote based gradient-decoding at the edge server.} This novel design will allow implementing AirComp across devices that are endowed with digital modulation chips, without requiring significant changes in the hardware or the communication architecture. 

Moreover, while existing works on over-the-air analog aggregation mostly rely on numerical experiments for performance evaluation  \cite{zhu2018low,amiri2019machine}, one of the main contributions of this work is an analytical study of the convergence rate of the OBDA scheme. The considered (model) convergence rate is defined as the rate at which { the expected value of average gradient norm}, denoted by $\bar{G}$, diminishes as the number of rounds, denoted by $N$, and the number of devices, denoted by $K$, grow. 
{\color{black}
For ideal FEEL with single-bit gradient quantization transmitted over noiseless channels, the convergence rate is shown to be \cite{2018signsgd}
\begin{align}\label{eq:noiseless}
\bar{G}\leq \frac{1}{\sqrt{N}}\left(c + \frac{c'}{\sqrt{K}}\right),
\end{align}
where $c$ and $c'$ denote some constants related to the landscape of the training loss function. 

The convergence rates of OBDA in the presence of channel hostilities are derived for three scenarios: 1) Gaussian MAC, 2) fading MAC with perfect (transmit) CSI, and 3) fading MAC with imperfect (transmit) CSI. For all three scenarios, the derived convergence rates share the following form:
\begin{align}\label{eq:channel_hostilities}
\bar{G} \leq \frac{a}{\sqrt{N}}\left(c + \frac{c'}{\sqrt{K}} + b\right),
\end{align}
where the channel hostilities are translated into a scaling factor $a\in (1, \infty)$ and a bias term $b\in (0, \infty)$. It is clear that the wireless channel imperfections slow down the model convergence with respect to the noiseless counterpart in \eqref{eq:noiseless}. The two terms $a$ and $b$ are independent of $N$, but are functions of $K$ and the received \emph{signal-to-noise ratio} (SNR) as shown explicitly in the sequel.
Particularly, as $K$ or the received SNR  grows,  $a$ and $b$ are shown to converge to their noiseless limits of 1 and 0, respectively. The convergence is shown to follow different scaling laws described as follows.

\begin{itemize}
\item \textbf{Gaussian MAC}: We consider a MAC with unit channel gain and additive complex Gaussian noise. For this case, the scaling factor $a$ and the bias term $b$ in \eqref{eq:channel_hostilities} scale as $1+ O\left(\frac{1}{K\sqrt{\text{SNR}}}\right)$ and $O\left(\frac{1}{K\sqrt{\text{SNR}}}\right)$, respectively.

\item \textbf{MAC with fading and perfect CSI}: 
{We consider a fading MAC and OBDA with perfect CSI, which is needed for transmit power control. Let $\alpha$ denote the expected ratio of gradient coefficients truncated due to the adopted truncated channel inversion power control under a power constraint. Then, the terms $a$ and $b$ in \eqref{eq:channel_hostilities} scale as $1 + O\left(\frac{1}{\alpha K\sqrt{\text{SNR}}}\right)$ and $O\left(\frac{1}{\alpha K\sqrt{\text{SNR}}}\right)$, respectively. In other words, the variation in channel quality due to fading is translated into an effective reduction on the number of devices by a factor of $\alpha$ in each round. This leads to a decreased  convergence rate compared to the case of Gaussian channels.}

\item \textbf{MAC with fading and imperfect CSI}: In practice, CSI errors may exist due to inaccurate channel estimation. Using imperfect CSI for power control results in  additional perturbation to the received gradients. As a result, the  terms $a$ and $b$ in \eqref{eq:channel_hostilities} scale as $1 + O\left(\frac{1}{\alpha\sqrt{K}}\right)$  and $O\left(\frac{1}{\alpha\sqrt{K}}\right)$, respectively, and are asymptotically independent of the receive SNR. Compared with the preceding two cases, the much slower rates and their insensitivity to increasing SNR shows that CSI errors can incur severe degradation of the OBDA performance. 

\end{itemize}
}

{
Finally, it is worth emphasizing that the current work differs from the closely related work  \cite{2018signsgd,reisizadeh2019fedpaq} in that our focus is on designing customized wireless communication techniques for implementing FEEL over a wireless network rather than developing new federated learning algorithms. To this end, we particularly take into account the characteristics of wireless channels in the design of communication techniques for FEEL, and characterize their effects on the resultant convergence rate. The key findings are summarized above, which are originally presented in this work and represent the novelty \emph{with respect to} (w.r.t.) \cite{2018signsgd,reisizadeh2019fedpaq}.}

\emph{Organization}: The remainder of the paper is organized as follows. Section II introduces the learning and communication models. Section III presents the proposed  OBDA scheme. The convergence analysis under the AWGN channel case is presented 
 in Section IV, and further extended to the fading channel case in Section V accounting for both the perfect CSI and imperfect CSI scenarios. Section VI presents the experimental results using real datasets followed by concluding remarks in Section VII.

\begin{figure*}[tt]
\centering
\includegraphics[width=14cm]{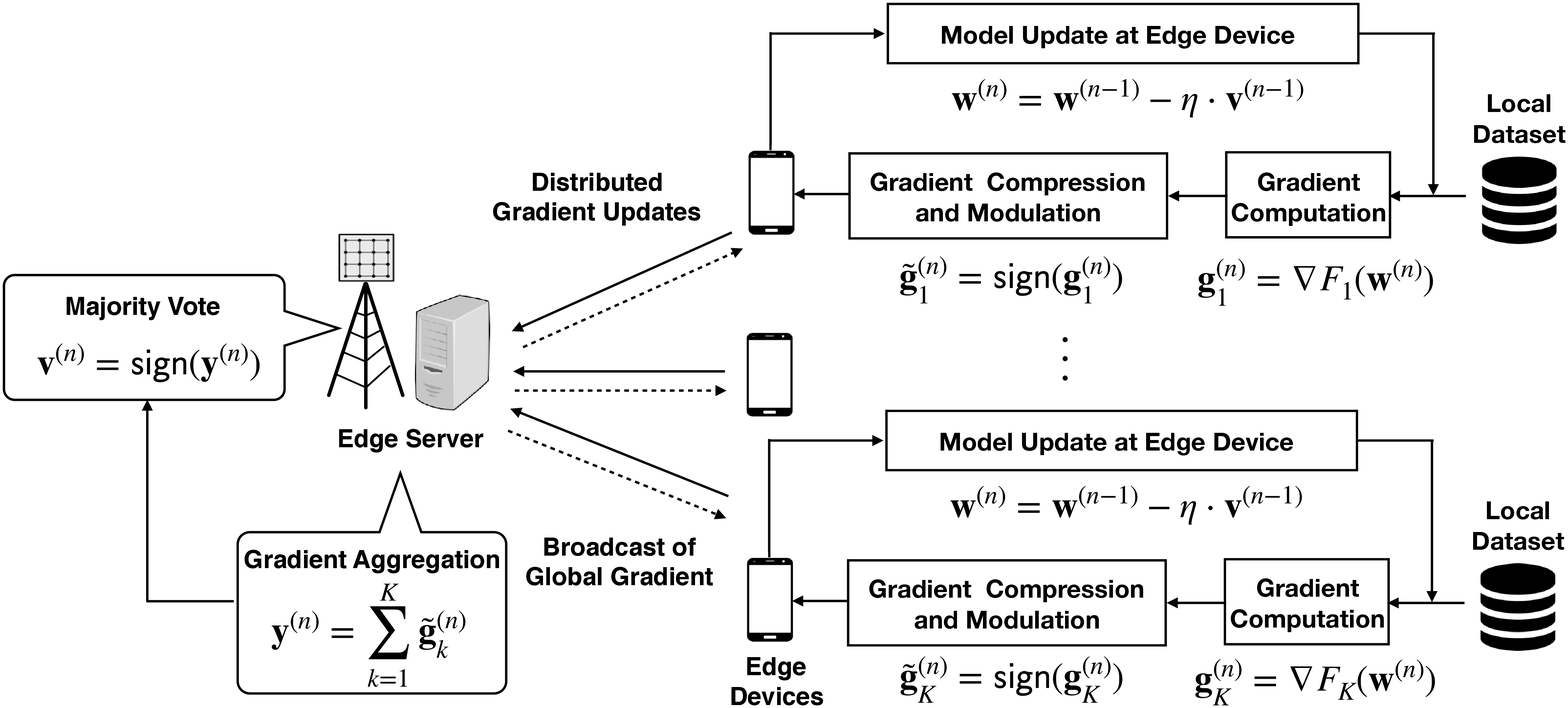}
\caption{FEEL via OBDA from distributed data.}
\vspace{-3mm}
\label{Fig:1}
\end{figure*}

\vspace{-1mm}
\section{Learning and Communication Models}
\vspace{-1mm}
We consider a FEEL system comprising a single edge server coordinating the learning process across
  $K$ edge devices as  shown in Fig.~\ref{Fig:1}. 
Device $k$, $k = 1,\ldots, K$, has its own local dataset, denoted by ${\cal D}_k$, consisting of labeled data samples $\{(\bx_i,y_j)\} \in {\cal D}_k$, where $\bx_i \in \mathbb{R}^d$ denotes the unlabeled data and $y_j \in \mathbb{R}$ the associated label. 
 A common model (e.g., a classifier), represented by the parameter vector $\bw \in \mathbb{R}^q$, is trained collaboratively across the edge devices, orchestrated by the edge server. Here $q$ denotes the model size. 

\subsection{Learning Model}
The loss function measuring the model error is defined as follows.
 The \emph{local loss function} of the model $\bw$ on ${\cal D}_k$ is
\begin{align}\label{eq:local_loss} \qquad
F_k(\bw) = \frac{1}{|{\cal D}_k|} \sum_{(\bx_j, y_j) \in {\cal D}_k} f(\bw, \bx_j, y_j),
\end{align}
where $f(\bw, \bx_j, y_j)$ denotes the sample loss quantifying the prediction error of the model $\bw$ on the training sample $\bx_j$  w.r.t. its true label $y_j$. 
For convenience, we rewrite $f(\bw, \bx_j, y_j)$ as $f_j(\bw)$, and  assume uniform sizes for local datasets; that is, $|{\cal D}_k| = D$, $\forall k$. Then, the \emph{global loss function} on all the distributed datasets can be written as
{
\begin{align}\label{eq:global_loss} \;\;
F(\bw) \triangleq \frac{\sum_{k=1}^K \sum_{j \in {\cal D}_k }f_j(\bw)}{\sum_{k=1}^K |{\cal D}_k| } =
 \frac{1}{K} \sum_{k = 1}^K F_k(\bw).
\end{align}
}

{
The goal of the learning process is thus to minimize the global loss function $F(\bw)$.\footnote{\color{black}
The problem formulation follows the standard  in the existing federated learning literature (see e.g., [3]-[22]). Note that the formulated problem for model training is a stochastic problem as we construct the loss function using a randomly sampled subset of data, and the minimization of the loss function is solved by \emph{stochastic gradient decent} (SGD). The statistical distribution of the stochastic gradient estimate is assumed to satisfy Assumption 3, which is a widely-adopted assumption for tractable convergence analysis for non-convex loss functions. Note that the SGD approach can also handle online training scenarios where the  dataset is collected sequentially in real-time.}
One way to do this is to  upload all the local datasets to the edge server, and solve the problem in a centralized manner. However, this is typically undesirable due to either privacy concerns, or the sheer size of the datasets.   
Alternatively, the FEEL framework can be employed to minimize $F(\bw)$  in a distributed manner. We focus on the gradient-averaging implementation of FEEL in the current work as illustrated in Fig. \ref{Fig:1} with the detailed procedure eloborated in the sequel. }

In each communication round of FEEL, say the $n$-th round, device $k$ computes a local estimate of the gradient of the loss function in \eqref{eq:global_loss} using its local dataset ${\cal D}_k$ and the current parameter-vector $\bw^{(n)}$.  Let $\bg_k^{(n)} \in \mathbb{R}^q$ denote the local gradient estimate at device $k$ in the $n$-th round, where we remove the dependence on the parameter vector $\bw^{(n)}$ for simplicity. We then have:
 \begin{align}\label{eq:local_update} \qquad
\bg_k^{(n)} =  \frac{1}{n_b} \sum_{j \in \tilde{\cal D}_k} \nabla f_j(\bw^{(n)}),
\end{align}
where $\nabla$ represents the gradient operator.   $\tilde{\cal D}_k \subset {\cal D}_k$ is the selected data batch from the local dataset for computing the local gradient estimate, and $n_b$ is the batch size. Accordingly, $n_b = |{\cal D}_k|$ means that all the local dataset is used for gradient estimation. 
{
If the local gradient estimates can be reliably conveyed to the edge server, the global estimate of the gradient of the loss function in \eqref{eq:global_loss}  would be computed as follows:\footnote{ For the case with heterogenous dataset sizes at different devices, the global gradient estimate is a weighted average of the local ones, i.e., $\bg^{(n)} = \frac{\sum_{k=1}^K |{\cal D}_k| \bg_k^{(n)}}{\sum_{k=1}^K |{\cal D}_k| }$. The desired weighted aggregation of the local gradient estimate can also be attained by the proposed over-the-air aggregation with an additional pre-processing $\phi_k(\cdot)$ on each of the transmitted signals, $x_k$, via $\phi_k(x_k) = \frac{|{\cal D}_k|}{\sum_{k=1}^K |{\cal D}_k| } x_k$ similarly as in \cite{yang2018federated}.}
 \begin{align}\label{eq:gradient_averaging} \qquad
\hat \bg^{(n)} = \frac{1}{K} \sum_{k=1}^K \bg_k^{(n)}.
\end{align} 
}
Then, the global gradient estimate is broadcast back to  each device, which then uses it to update the current model via gradient descent based on the following equation:
 \begin{align}\label{eq:model_update} \qquad
\bw^{(n+1)} = \bw^{(n)} - \eta \cdot \hat \bg^{(n)},
\end{align}
where $\eta$ denotes the learning rate. 
The steps in \eqref{eq:local_update}, \eqref{eq:gradient_averaging}, and \eqref{eq:model_update} are iterated until a convergence condition is met. 



{
As observed from \eqref{eq:gradient_averaging}, it is only the aggregated gradient, i.e., $\sum_{k=1}^K \bg_k$,
not the individual gradient estimates $\{\bg_k\}$, needed at the edge server for computing the global gradient estimate.}
 This motivates the communication efficient aggregation scheme  presented in Section \ref{sec:OBDA Design}.

\vspace{-1mm}
\subsection{Communication Model} 
\vspace{-1mm}
Local gradient estimates of edge devices  are transmitted to the edge server over a broadband MAC. To cope with the frequency selective fading and inter-symbol interference, OFDM modulation is adopted to divide the available bandwidth $B$ into $M$ orthogonal sub-channels. 
We assume that a fixed digital constellation is employed by all the devices to transmit over each sub-channel. Thus, each device needs to transmit its local gradient estimate using a finite number of digital symbols. This requires quantization of the local gradient estimates, and mapping each quantized gradient element to one digital symbol to facilitate the proposed OBDA. Let $\tilde \bg_k = [\tilde g_{k,1}, \ldots, \tilde g_{k,q}]^T$ denote the channel input vector  of the $k$-th device, where $q$ is the size of the gradient vector, and $\tilde \bg_{k,j} \in {\cal Q}$ for some finite digital input constellation ${\cal Q}$.



During the gradient-uploading phase, all the devices transmit simultaneously over the whole available bandwidth.
 At each communication round, gradient-uploading duration consists of $N_s = \frac{q}{M}$ OFDM symbols for complete uploading of all the gradient parameters. We assume symbol-level synchronization among the transmitted devices through a synchronization channel (e.g., ``timing advance'' in LTE systems \cite{TALTE}).\footnote{The accuracy of  synchronization is proportional to the bandwidth dedicated for the synchronization channel. Particularly, the current state-of-the-art phase-locked loop can achieve a synchronization offset of $0.1B_s^{-1}$,  where  $B_s$  is the amount of bandwidth used for synchronization. In  existing LTE systems, the typical value of $B_s$ is 1MHz. Thus, a sufficiently small synchronization offset of  $0.1\mu sec$ can be achieved. Note that in a broadband OFDM system, as long as the synchronization offset is smaller than the length of cyclic prefix (the typical value is $5 \mu sec$ in the LTE systems), the offset simply introduces a phase shift to the received symbol. The phase shift can be easily compensated by channel equalization, incurring no performance loss \cite{arunabha2010fundamentals}.} 
 Accordingly, the $i$-th aggregated gradient parameter, denoted by $\tilde g_i$, with $i = (t-1)M + m$, received at the $m$-th sub-carrier and $t$-th OFDM symbol is given by
 \begin{align}\label{channel_model}  \qquad
  \vspace{-1mm}
 \tilde g_i = \sum_{k=1}^K  h_k[t,m] p_k[t,m] \tilde g_{k,i} + z[t,m], \qquad \forall i,
   \vspace{-1mm}
 \end{align}
{\color{black} where  $\{h_k[t,m]\}$ are the channel coefficients with identical Rayleigh distribution, i.e.,  $h_k[t, m] \sim {\cal CN}(0,1)$;}\footnote{\color{black} It should be emphasized that the current analysis does not require the assumption of independent channel realizations over different time slots. In particular, the same analytical results hold as long as the channel coefficients at different time slots follow identical Rayleigh (complex Gaussian) distribution, regardless of whether there is correlation over time or not.} \footnote{ We also assume identical path-losses for different devices to simplify exposition. Note that the difference in path-losses between devices can be equalized by the channel inversion power control applied in the design as specified in the sequel. If there exists a bottleneck device with severe path-loss that does not allow channel inversion within the given power budget, it can be excluded in the scheduling phase.}
and $p_k[t,m]$ is the associated power control policy  to be specified in the sequel. Finally,  $z[t,m]$ models the zero-mean i.i.d. \emph{additive white Gaussian noise} (AWGN) with variance $\sigma_z^2$.
For ease of notation, we skip the index of OFDM symbol $t$ in the subsequent exposition whenever no confusion is incurred.


The power allocation over sub-channels, $\{p_k[m]\}$, will be adapted to the corresponding channel coefficients, $\{h_k[m]\}$, for implementing gradient aggregation via AirComp as presented in the sequel. The transmission of each device is subject to a long-term transmission power constraint:
\begin{align}\label{power_constraint1}
{\mathbb E} \l[{ \sum_{m=1}^M |p_k[m]} |^2  \r] \leq P_0, \qquad \forall k,
\end{align}
where the  expectation is taken over the distribution of random channel coefficients, and we assume ${\mathbb E} \l[ \tilde g_{k,i} \tilde g_{k,i}^* \r] = 1$ without loss of generality. 
Since channel coefficients are identically distributed over different sub-channels, the above power constraint reduces to
\begin{align}\label{power_constraint2}   \qquad
{\mathbb E} \l[  | p_k[m]|^2 \r]  &\leq \frac{P_0}{M}, \qquad \forall k.
\end{align}

%

\vspace{-3mm}
\section{One-Bit Broadband Digital Aggregation (OBDA): System Design}\label{sec:OBDA Design}
\vspace{-1mm}
As discussed, the essential idea of OBDA is to integrate signSGD and AirComp so as to support communication-efficient FEEL using digital modulation. The implementation of the idea requires an elaborate system design, which is explained in detail in this section.   

\subsection{Transmitter Design}
\begin{figure*}[tt]
 \centering
   \subfigure[Transmitter design for edge devices]{\label{subfig:tx}\includegraphics[width=0.8\textwidth]{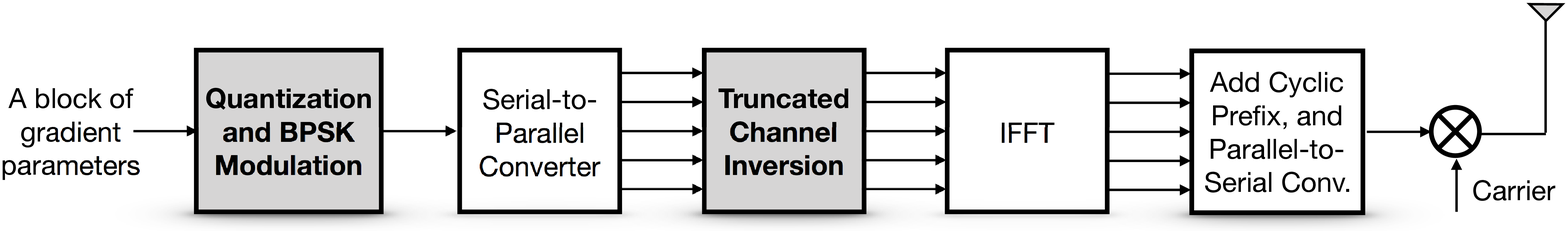}}
  \hspace{0.25in}
  \subfigure[Receiver design for edge server]{\label{subfig:rx}\includegraphics[width=0.8\textwidth]{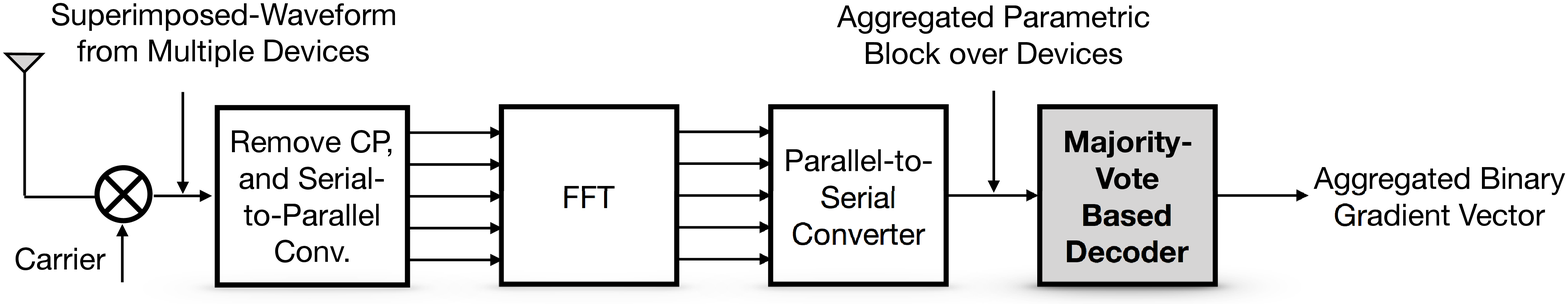}}
   \vspace{3mm}
  \caption{Transceiver design for OBDA.}
  \label{Fig:Fig:system_diagram}
  \vspace{-3mm}
\end{figure*}

The transmitter design for edge devices is shown  in  Fig. \ref{subfig:tx}.  The design builds on a conventional OFDM transmitter with \emph{truncated channel-inversion power control}. However, unlike in conventional communication systems, where coded data bits are passed to the OFDM encoder, here we feed raw quantized bits without any coding. 
Inspired by signSGD \cite{2018signsgd}, we apply \emph{one-bit quantization} of local gradient estimates, which simply corresponds to taking the signs of the local gradient parameters element-wise:
\begin{align} (\text{One-bit quantization}) \qquad
\tilde g_{k,i} = {\sf sign}(g_{k,i}),    \quad \forall k,i.
\end{align}
Each of the binary gradient parameters is modulated into one \emph{binary phase shift keying} (BPSK) symbol. We emphasize that, even though we use BPSK modulation in our presentation and the convergence analysis for simplicity, the  extension of OBDA to 4-QAM configuration is straightforward by simply viewing each 4-QAM symbol as two orthogonal BPSK symbols. Indeed, we employ  4-QAM modulation for the numerical experiments in Section \ref{simulation}.
 The long symbol sequence is then divided into blocks, and each block of $M$ symbols is transmitted as a single OFDM symbol with one symbol/parameter over each frequency sub-channel. 

Assuming perfect CSI at the transmitter, sub-channels are inverted by power control so that gradient parameters transmitted by different devices are received with identical amplitudes, achieving amplitude alignment at the receiver as required for OBDA.  Nevertheless, a brute-force approach is inefficient if not impossible under a power constraint since some sub-channels are likely to encounter deep fades. To overcome the issue, we adopt the more practical \emph{truncated-channel-inversion} scheme \cite{zhu2018low}. To be specific, a sub-channel is inverted only if its gain exceeds a so called \emph{power-cutoff threshold}, denoted by $g_{\sf th}$, or otherwise allocated zero power. Then the transmission power of the $k$-th device on the $m$-th sub-channel, $p_k[m]$, is 
\begin{align}\label{truncated_channel_inv} \qquad
 p_k[m] = \l\{\begin{aligned} 
& \frac{\sqrt{\rho_0}}{ |h_k[m]|} \frac{h_k[m]^\dagger}{|h_k[m]|}, && |h_k[m]|^2 \geq g_{\sf th} \\
&0, &&  |h_k[m]|^2 < g_{\sf th},
\end{aligned}
\r.
 \end{align}
where $\rho_0$ is a scaling factor set to satisfy the average-transmit-power constraint in \eqref{power_constraint2}, and determines the receive power of the gradient update from each device as observed from \eqref{channel_model}. The exact value of $\rho_0$ can be computed via
\begin{align}
\rho_0 = \frac{P_0}{M{\sf Ei}(g_{\sf th})},
\end{align}
where ${\sf Ei}(x)$ is the exponential integral function defined as  ${\sf Ei}(x) = \int_x^\infty \frac{1}{t} \exp(-t) dt$. The result follows from the fact that the channel coefficient is Rayleigh distributed $h_k[m] \sim {CN(0,1)}$, and thus the channel gain $g_k = |h_k[m]|^2$ follows an exponential distribution with unit mean. Thus the power constraint in \eqref{power_constraint2} is explicitly given by
$\rho_0 \int_{g_{\sf th}}^\infty \frac{1}{g}\exp(-g) dg = \frac{P_0}{M}$. The desired result follows by solving the integral. 

We remark that the policy can cause the loss of those gradient parameters that are mapped to the truncated sub-channels. To measure the loss, we define the \emph{non-truncation probability} of a parameter, denoted by  $\alpha$, as the probability that the associated channel gain is above the power-cutoff threshold:
\begin{align}\label{eq:truncation_ratio}
\alpha = {\sf Pr}(|h_k|^2 \geq g_{\sf th}) = \exp(-g_{\sf th}).
\end{align}
The result immediately follows from the exponential distribution of the channel gain.
The value of $\alpha$ affects the learning convergence rate as shown in the sequel. 

\vspace{-4mm}
\subsection{Receiver Design}
\vspace{-1mm}
Fig. \ref{subfig:rx} shows the receiver design for the edge server. It has the same architecture as a conventional OFDM receiver
except that the digital detector is replaced with a \emph{majority-vote based decoder} for estimating the global gradient-update from the received signal as elaborated in the following. 

Consider an arbitrary communication round. Given the simultaneous transmission of all participating devices, the server receives superimposed waveforms. By substituting the truncated-channel-inversion policy in \eqref{truncated_channel_inv} into \eqref{channel_model}, the server obtains the aggregated local-gradient block, denoted by a $M\times 1$ vector $\tilde \bg[t]$,  at the parallel-to-serial converter output [see Fig. \ref{subfig:rx}] as: 
\begin{align}\label{channel_model2} \!\! (\text{Over-the-air aggregation}) \;
\tilde \bg[t] = \sum_{k =1}^K \sqrt{\rho_0} \tilde \bg^{(\sf Tr)}_{k}[t] + \bz[t], \!\!
\end{align}
where $t$ is the index of the local-gradient block (OFDM symbol) as defined in \eqref{channel_model}, $\tilde \bg^{(\sf Tr)}_{k}[t]$ is the truncated version of $\tilde \bg_{k}[t] = [\tilde g_{k, (t-1)M + 1}, \ldots, \tilde g_{k,tM}]^T$, with the truncated elements determined by the channel realizations according to \eqref{truncated_channel_inv} and set to zero. 
Next, cascading all the $N_s$ blocks recovers the full-dimension aggregated one-bit quantized local gradient estimates:
\begin{align}\label{eq:aggregated_gradient}  \qquad
\tilde \bg = \l[\tilde \bg[1]^T, \tilde \bg[2]^T, \ldots, \tilde \bg[N_s]^T\r]^T.
\end{align}
Finally, to attain a global gradient estimate from $\tilde \bg$  for model updating, a majority-vote based decoder is adopted and enforced by simply taking the element-wise sign of $\tilde \bg$:
\begin{align} (\text{Majority-vote based decoder}) \qquad
\bv = {\sf sign} (\tilde \bg).
\end{align}
The operation essentially estimates the global gradient-update by  over-the-air element-wise majority vote based on the one-bit quantized local gradient estimates attained at different devices.

{
\begin{remark}[Why majority vote?]
\emph{ Instead of simply taking an average of the one-bit local gradient estimates, majority vote that involves additional sign-taking operation brings additional benefit from the following two aspects: 1) it is more hardware-friendly and robust (against noise) to detect the sign of the aggregated gradient estimate (majority vote) than recovering its actual value in full precision at the edge server;  2) it allows  communication-efficient broadcasting of the aggregated gradient estimates back to the edge devices thanks to the one-bit quantized elements after the majority vote.
}
\end{remark}
}

 Then, the server initiates the next communication round by broadcasting the global gradient estimate to all the devices for model updating via 
 \vspace{-3mm}
\begin{align}\label{eq:model_update_onebit} \qquad
\vspace{-2mm}
\bw^{(n+1)} = \bw^{(n)} - \eta \bv^{(n)},
\end{align}
or completes the learning process if the convergence criterion (e.g., target number of communication rounds) is met.  {We assume that the global gradient parameters can be sent to the devices perfectly, due to the high transmit power available at the edge server and the use of the whole downlink bandwidth for broadcasting.}


\section{Convergence Analysis for OBDA over AWGN Channels}\label{sec: AWGN_analysis}
In this section, we formally characterize the learning performance of a FEEL system deploying the  proposed OBDA scheme in Section \ref{sec:OBDA Design} over static AWGN channels; that is, we assume $h_k[t,m] = 1$, $\forall k,t,m$, in this section.  Particularly, we focus on understanding how the channel noise affects the convergence rate of the proposed scheme. 

\vspace{-3mm}
\subsection{Basic Assumptions}\label{sec:assumptions}
\vspace{-1mm}
To facilitate the convergence analysis, several standard assumptions are made on the loss function and computed gradient estimates. In order to allow the developed theory to be applicable to the popular \emph{deep neural networks} (DNNs), we do not assume a convex loss function, but require a lower bounded one as formally stated below, which is the minimal assumption needed for ensuring convergence to a stationary point \cite{AllenZhu2017natasha2}.

\begin{assumption}[Bounded Loss Function] \emph{For any parameter vector $\bw$, the associated loss function is lower bounded by some value $F^*$}, i.e., $F(\bw) \geq F^*$, $\forall \bw$.
\end{assumption}

\noindent Assumptions 2 and 3 below, on the Lipschitz smoothness and bounded variance, respectively, are standard in the stochastic optimization literature \cite{AllenZhu2017natasha2}.  

\begin{assumption}[Smoothness] \emph{Let  $\bg$  denote the gradient of the loss function $F(\bw)$ in \eqref{eq:global_loss} evaluated at point $\bw = [w_1,\cdots, w_q]$ with $q$ denoting the number of model parameters. We assume  that there exists a vector of non-negative constants ${\bf L} = [L_1, \cdots, L_q]$, for any $\bw, \bw'$, that satisfy}
\begin{align}\label{eq:smoothness}
\!\!\! |F(\bw') - [F(\bw) + \bg^T(\bw' - \bw)]| \leq \frac{1}{2} \sum_{i=1}^q L_i(w'_i - w_i)^2. \!\!\!
\end{align}
\end{assumption}

\begin{assumption}[Variance Bound] \emph{It is assumed that the stochastic gradient estimates $\{\bg_j\}$ defined in \eqref{eq:local_update} are independent and unbiased estimates of the batch gradient $\bg = \nabla F(\bw)$ with  coordinate bounded variance, i.e.,}
\begin{align}
&\mathbb{E}[\bg_j] = \bg, \qquad \forall j, \\
&\mathbb{E}[(g_{j,i} - g_i)^2] \leq \sigma_i^2 \qquad \forall j, i,
\end{align}
\emph{where $g_{j,i}$ and $g_i$ denote the $i$-th element of $\bg_j(\bw)$ and $\bg(\bw)$, respectively, and $\boldsymbol \sigma = [\sigma_1, \ldots, \sigma_q]$ is a vector of non-negative constants.}
\end{assumption}

We further assume that the data-stochasticity induced gradient noise, which causes the discrepancy between $\bg_j$ and $\bg$, is unimodal and symmetric, as verified by experiments in \cite{2018signsgd} and formally stated below.

\begin{assumption}[Unimodal, Symmetric Gradient Noise]\emph{For any given parameter vector $\bw$, each element of the stochastic gradient vector $\bg_j(\bw)$, $\forall j$, has a unimodal distribution that is also symmetric around its mean (the ground-truth full-batch gradient elements).}
\end{assumption}
\noindent Clearly, Gaussian noise is a special case. Note that even for a moderate mini-batch size, we expect the central limit theorem to take effect and render typical gradient noise distributions close to Gaussian. 

\vspace{-1mm}
\subsection{Convergence Analysis}
\vspace{-1mm}
The above assumptions allow tractable convergence analysis as follows. 
Given AWGN channels, the gradient aggregation is a direct consequence of the MAC output and the power control in \eqref{truncated_channel_inv} is not needed in this case.  
Specifically, without truncation due to fading, the full-dimension aggregated local gradient defined in \eqref{eq:aggregated_gradient} is given by
\begin{align}\label{eq:aggregated_gradient_AWGN}
\tilde \bg = \sum_{k =1}^K \sqrt{\rho_0} \tilde \bg_{k} + \bz, 
\end{align}
where we have $\rho_0 = \frac{P_0}{M}$ according to the power constraint in \eqref{power_constraint2}.

{
The resulting convergence rate of the proposed OBDA scheme is derived as follows. Throughout the paper, we set the learning rate to $\eta = \frac{1}{\sqrt{\|{\bf L}\|_1 n_b}}$ and the mini-batch size to $n_b = \frac{1}{\gamma}N$, with an arbitrary $\gamma > 0$ and $N$ denoting the number of communication rounds. }

{
\begin{theorem}\label{theo:AWGN}
\emph{Consider a FEEL system deploying OBDA over AWGN channels, the  convergence rate is given by}
\begin{multline}
\!\! {\mathbb E} \left[\frac{1}{N} \sum_{n=0}^{N-1} \|{\bf g}^{(n)}\|_1\right] \leq  \frac{a_{\sf AWGN}}{\sqrt{N}}\left(\sqrt{\|{\bf L}\|_1} (F^{(0)} - F^* + \frac{\gamma}{2}) + \right. \\
\left. \frac{2\gamma}{\sqrt{K}}\|\boldsymbol{\sigma}\|_1 + {b_{\sf AWGN} } \right), 
\end{multline}
\emph{where the scaling factor $a_{\sf AWGN}$ and the bias term $b_{\sf AWGN}$ are given by}
\begin{align}
a_{\sf AWGN} = \frac{1}{1- \frac{1}{K \sqrt{\rho}}}, \qquad
{b_{\sf AWGN} = \frac{2 \gamma}{K\sqrt{\rho} } \|\boldsymbol{\sigma}\|_1 },
\end{align}
\emph{and $\rho \triangleq \frac{\rho_0}{\sigma_z^2} = \frac{P_0}{M \sigma_z^2}$ denotes the \emph{receive SNR}. 
}
\end{theorem}
}
\proof See Appendix \ref{app:theo:AWGN}.
\endproof


For comparison, we reproduce below the convergence rate over noiseless channels derived as Theorem 2 in \cite{2018signsgd}.
{
\begin{lemma}\label{lemma:error_free}
\emph{The convergence rate with OBDA over error-free channels is given by}
\begin{multline}
{\mathbb E} \left[\frac{1}{N} \sum_{n=0}^{N-1} \|{\bf g}^{(n)}\|_1\right] \leq \frac{1}{\sqrt{N}}\left(\sqrt{\|{\bf L}\|_1} (F^{(0)} - F^* + \frac{\gamma}{2})  + \right. \\
\left. \frac{2\gamma}{\sqrt{K}}\|\boldsymbol{\sigma}\|_1 \right).
\end{multline}
\end{lemma}
}

\vspace{-3mm}
\begin{remark}[Effect of Channel Noise]
\emph{
A comparison between the results in Theorem \ref{theo:AWGN} and Lemma \ref{lemma:error_free} reveals that the existence of channel noise slows down the convergence rate by adding a scaling factor and a positive bias term, i.e., $a_{\sf AWGN}$ and $b_{\sf AWGN}$, respectively, to the upper bound on the time-averaged gradient norm. Due to the increased bound, more communication rounds will be needed for convergence. Nevertheless, the negative effect of channel noise vanishes at a scaling rate of $\frac{1}{K}$ as the number of participating devices grows. We can also see that we recover the convergence rate in Lemma \ref{lemma:error_free}  when $\rho \to \infty$.
}
\end{remark}
\section{Convergence Analysis for OBDA over Fading Channels}\label{sec: fading_analysis_perfect_CSI}
In this section, we extend the  convergence result for AWGN channels  to the more general case of fading channels. 
For this case, transmit CSI 
 is needed for power control. We consider both the cases of perfect CSI and imperfect CSI in the analysis.
  The same set of assumptions as in Section \ref{sec:assumptions} are made here.

\vspace{-3mm}
\subsection{Convergence Rate with Perfect CSI}
With perfect CSI at each device, the truncated channel inversion power control can be accurately performed. The resultant convergence rate of the OBDA scheme is derived as follows.
{
\begin{theorem}\label{theo:fading}
\emph{Consider a FEEL system deploying OBDA over fading channels with truncated channel inversion power control using perfect CSI, the  convergence rate is given by}
\begin{multline}
{\mathbb E} \left[\frac{1}{N} \sum_{n=0}^{N-1} \|{\bf g}^{(n)}\|_1\right] \leq  \frac{a_{\sf FAD}}{\sqrt{N}}\left(\sqrt{\|{\bf L}\|_1} (F^{(0)} - F^* + \frac{\gamma}{2}) + \right. \\ 
\left. \frac{2\gamma}{\sqrt{K}}\|\boldsymbol{\sigma}\|_1 + {b_{\sf FAD} } \right), 
\end{multline}
\emph{where the scaling factor $a_{\sf FAD}$ and the bias term $b_{\sf FAD}$ are}
\begin{align}
a_{\sf FAD} = \frac{1}{1- {(1-\alpha)^K} - \frac{2}{{\alpha K} \sqrt{\rho}}}, \;
b_{\sf FAD} = \frac{4\gamma}{\alpha K \sqrt{\rho}} \|\boldsymbol{\sigma}\|_1,
\end{align}
\emph{and $\rho \triangleq \frac{\rho_0}{\sigma_z^2} = \frac{P_0}{M {\sf Ei}(g_{\sf th}) \sigma_z^2}$ denotes the average \emph{receive SNR}. 
}
\end{theorem}
}
\proof See Appendix \ref{app:theo:fading}
\endproof

\begin{remark}[Effect of Channel Fading]
\emph{
A comparison between Theorems \ref{theo:AWGN} and \ref{theo:fading}  reveals that the existence of channel fading further slows down the convergence rate of the OBDA by introducing a larger scaling factor and a bias term:  $a_{\sf FAD} > a_{\sf AWGN}$ and $b_{\sf FAD} > b_{\sf AWGN}$. This negative effect of channel fading vanishes at a scaling rate of $\frac{1}{\alpha K}$ as the number of participating devices grows. 
Compared with the AWGN counterpart, the rate is slowed down by a factor of $\alpha$. The degradation is due to the gradient truncation induced by truncated channel inversion power control to cope with fading. 
}
\end{remark}

\vspace{-4mm}
\subsection{Convergence Rate with Imperfect CSI} \label{sec: fading_analysis_imperfect_CSI}
\vspace{-1mm}
 In practice, there may exist channel estimation errors that lead to imperfect channel inversion and, 
as a result, reduces the convergence rate.   
To facilitate the analysis, we adopt the bounded channel estimation error model (see e.g., \cite{hong2014signal}), where the estimated CSI is a perturbed version of the ground-true one and the additive perturbation, denoted as $\Delta$, is assumed to be bounded:
\begin{align}\label{eq:channel_estimation_error}
\hat h_k[m] = h_k[m] + \Delta, \qquad \forall k, m,
\end{align}
where we assume the absolute value of the perturbation is bounded by $|\Delta| \leq \Delta_{\max}\ll \sqrt{g_{\sf th}}$\footnote{When there exist channel estimation errors, it is desirable to set a relative high channel cutoff threshold $g_{\sf th}$ to ensure that $g_{\sf th} \gg \Delta_{\max}^2$ for avoiding the channel truncation decision misled by the estimation perturbation $\Delta$.}
 with a zero mean ${\mathbb E}(\Delta) = 0$, and a variance of ${\sf Var}(\Delta) = \sigma_\Delta^2$.
   
Based on the above CSI model, the model convergence rate can be derived below. 
{
\begin{theorem}\label{theo:imperfect_CSI}
\emph{Consider a FEEL system deploying OBDA over fading channels with truncated channel inversion power control using imperfect CSI, the convergence rate is given by
}
\begin{multline}
{\mathbb E} \left[\frac{1}{N} \sum_{n=0}^{N-1} \|{\bf g}^{(n)}\|_1\right] \leq  \frac{a_{\sf CERR}}{\sqrt{N}}\left(\sqrt{\|{\bf L}\|_1} (F^{(0)} - F^* + \frac{\gamma}{2}) + \right. \\
\left.  \frac{2\gamma}{\sqrt{K}}\|\boldsymbol{\sigma}\|_1 + {b_{\sf CERR} } \right), 
\end{multline}
\emph{where the scaling factor $a_{\sf CERR}$ and the bias term $b_{\sf CERR}$ are given by}
\begin{align}\label{eq:scaling_bias_imperfect_CSI}
 a_{\sf CERR} & = \frac{1}{1- {(1-\alpha)^K} - \frac{2}{{\alpha K \sqrt{\rho}}}-  \frac{2\sqrt{6} \sigma_\Delta}{\sqrt{\alpha K} \sqrt{\sqrt{g_{\sf th}} - \Delta_{\max}}}}, \notag\\
b_{\sf CERR} & = \l(\frac{4}{\alpha K \sqrt{\rho}}+{ \frac{4\sqrt{6} \sigma_\Delta}{\sqrt{\alpha K} \sqrt{\sqrt{g_{\sf th}}- \Delta_{\max}}}}\r) \gamma \|\boldsymbol{\sigma}\|_1,
\end{align}
\emph{and $\rho \triangleq \frac{\rho_0}{\sigma_z^2} = \frac{P_0}{M {\sf Ei}(g_{\sf th}) \sigma_z^2}$ denotes the average \emph{receive SNR}. 
}
\end{theorem}
}
\proof See Appendix \ref{app:theo:imperfect_CSI}
\endproof

\begin{remark}[Effect of Imperfect CSI]
\emph{
Comparing Theorem \ref{theo:imperfect_CSI} and Theorem \ref{theo:fading}, one can observe that the imperfect CSI reduces the convergence rate for OBDA  even further as reflected by larger scaling factor and bias terms: $a_{\sf CERR} > a_{\sf FAD}$ and $b_{\sf CERR} > b_{\sf FAD}$. Particularly, with imperfect CSI, the negative effect of channel fading vanishes at a slower scaling law of $\frac{1}{\sqrt{\alpha K}}$ as the number of participating devices increases. 
In contrast, it is a $\frac{1}{\alpha K}$ for the perfect CSI case, which is much faster. 
The results in \eqref{eq:scaling_bias_imperfect_CSI} also quantify the effect of the level of CSI accuracy (represented by $\Delta_{\max}$) on the convergence rate for the proposed scheme.
}
\end{remark}

\begin{figure*}[tt]
\centering
\includegraphics[width=14cm]{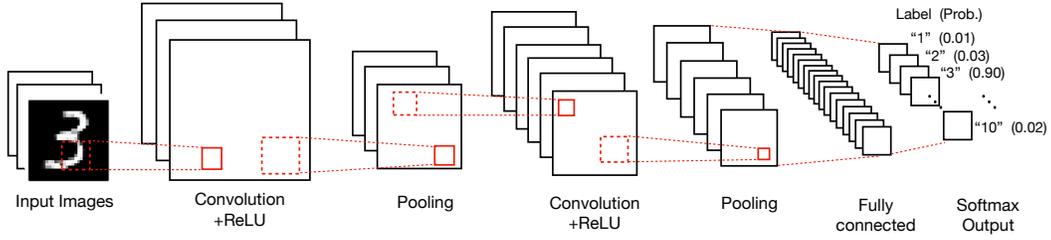}
\caption{Architecture of the CNN used in our experiments.}
\label{Fig:convolutional_neural_network}
\vspace{-4mm}
\end{figure*}

\section{Simulation results}\label{simulation}
For numerical experiments we consider a FEEL system with one edge server and $K = 100$ edge devices.
The simulation settings are given as follows unless specified otherwise.  The number of sub-channels is $M=1000$, and the average receive SNR, defined as $\rho = \frac{P_0}{M\sigma_z^2}$, is set to be 10 dB. { We consider the learning task of image classification using the well-known MNIST and CIFAR10 datasets, where the former consists of $10$ classes of black-and-white digits ranging from ``$0$" to ``$9$" and the latter comprises $10$ categories of colorful objectives such as airplanes, cars, etc.. In particular, for the MNIST dataset, as illustrated in Fig. \ref{Fig:convolutional_neural_network}, the classifier model is implemented using a $6$-layer \emph{convolutional neural network} (CNN)  that consists of two $5\times5$ convolution layers with ReLU activation (the first with $32$ channels, the second with $64$), each followed with a $2\times2$ max pooling; a fully connected layer with $512$ units,  ReLU activation; and a final softmax output layer (i.e., $q=582,026$). While for the CIFAR10 dataset, the well-known classifier model, ResNet18 with batch normalization proposed in \cite{he2016deep}, is applied.} We adopt the 4-QAM for the quantized gradient element modulation, where the odd and even gradient coefficients are mapped to the real and imaginary parts of 4-QAM symbols, respectively.  The open-source codes are available at https://github.com/BornHater/One-bit-over-the-air-computation.


\subsection{Performance Evaluation of OBDA}
\begin{figure}[tt]
\centering
\subfigure[Dataset: MNIST; Classifier: CNN ]{\includegraphics[width=0.42\textwidth]{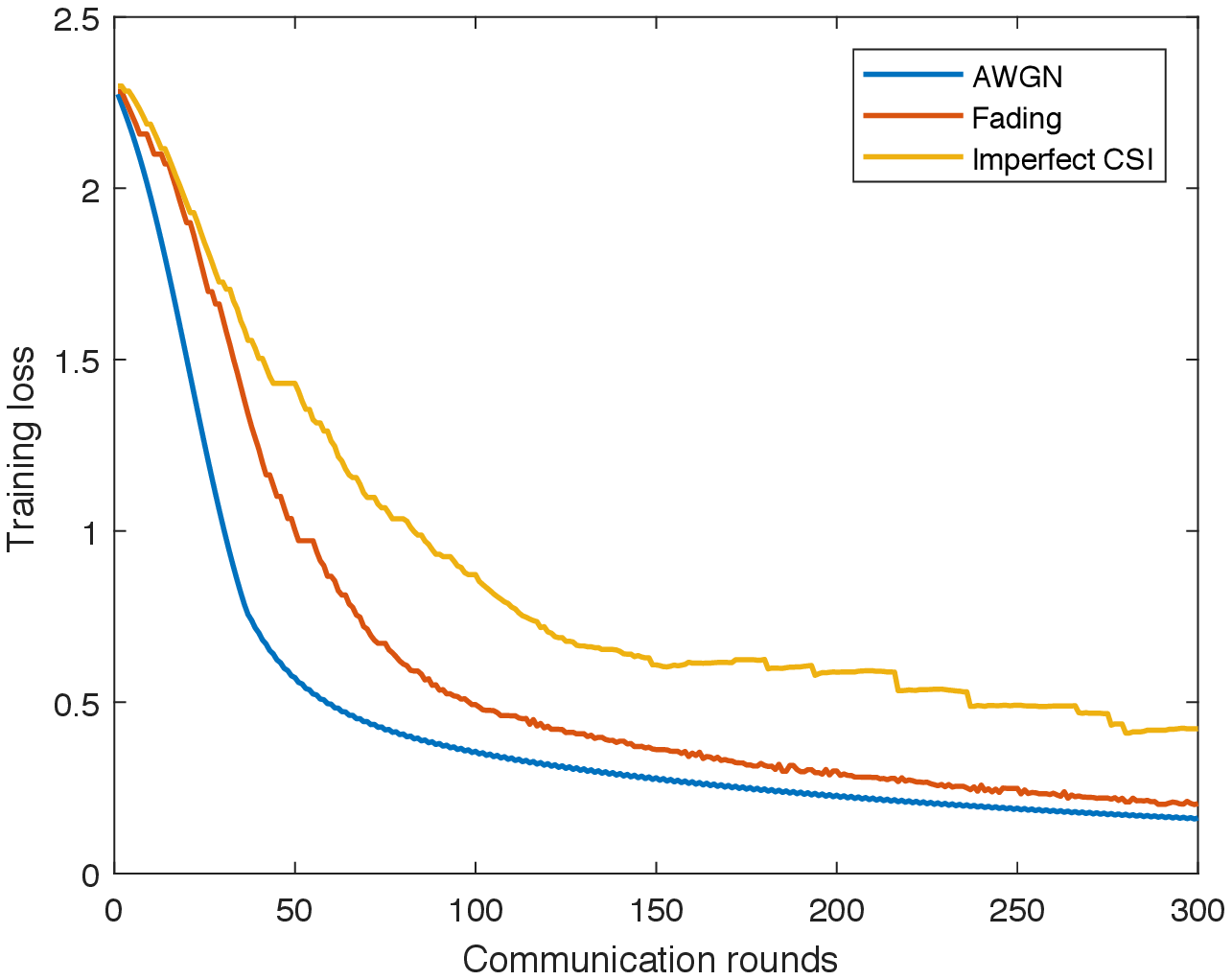}}
\subfigure[Dataset: CIFAR10; Classifier: ResNet18 ]{\includegraphics[width=0.42\textwidth]{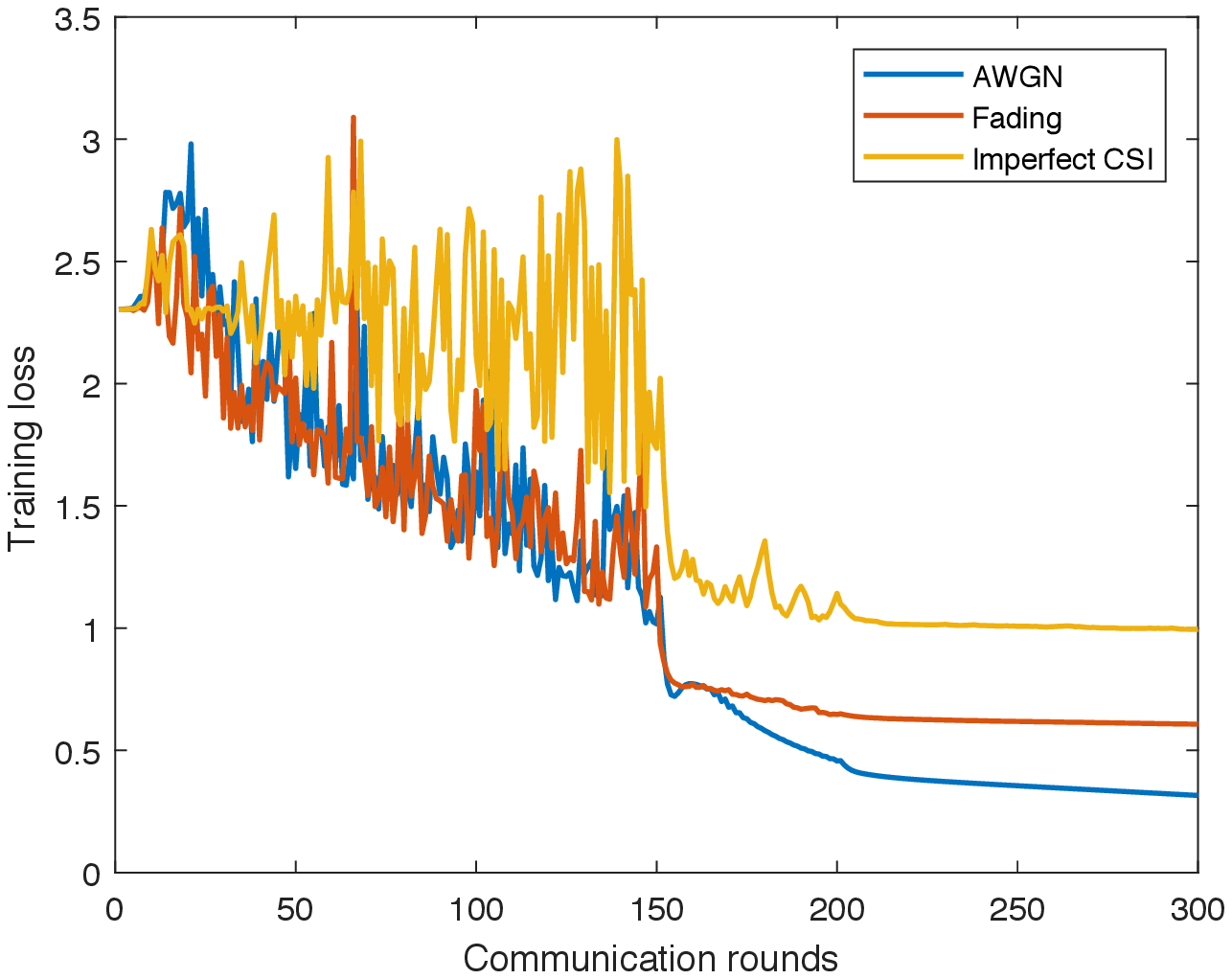}}
\vspace{3mm}
\caption{Convergence performance of FEEL using OBDA.
 }
\label{Fig:performacne_comparison}
\end{figure}

For both MNIST and CIFAR10 datasets, the effectiveness of the OBDA is evaluated in the three considered scenarios, namely over an AWGN MAC, and fading MACs with and without perfect CSI, which represent three levels of wireless hostilities. Test accuracy is plotted as a function of the number of communication rounds  in Fig.~\ref{Fig:performacne_comparison}.  The proposed OBDA scheme converges in all three scenarios, but at different rates depending on the level of wireless hostility the scheme suffers from. 
In the presence of channel fading the convergence is slower compared with its counterpart over an AWGN channel. This is because part of the gradient signs corresponding to subchannels experiencing deep fade are truncated, rendering a smaller number of effective participating devices for each gradient entry.  
The imperfect CSI further slows down the convergence of the proposed OBDA. This is due to inaccurate aggregation, which results in a deviated gradient for model updating.  These observations are aligned with our analysis presented  in Theorems \ref{theo:AWGN}-\ref{theo:imperfect_CSI}.


\subsection{Effect of  Device Population}
\begin{figure}[tt]
\centering
\subfigure[Dataset: MNIST; Classifier: CNN  ]{\includegraphics[width=0.42\textwidth]{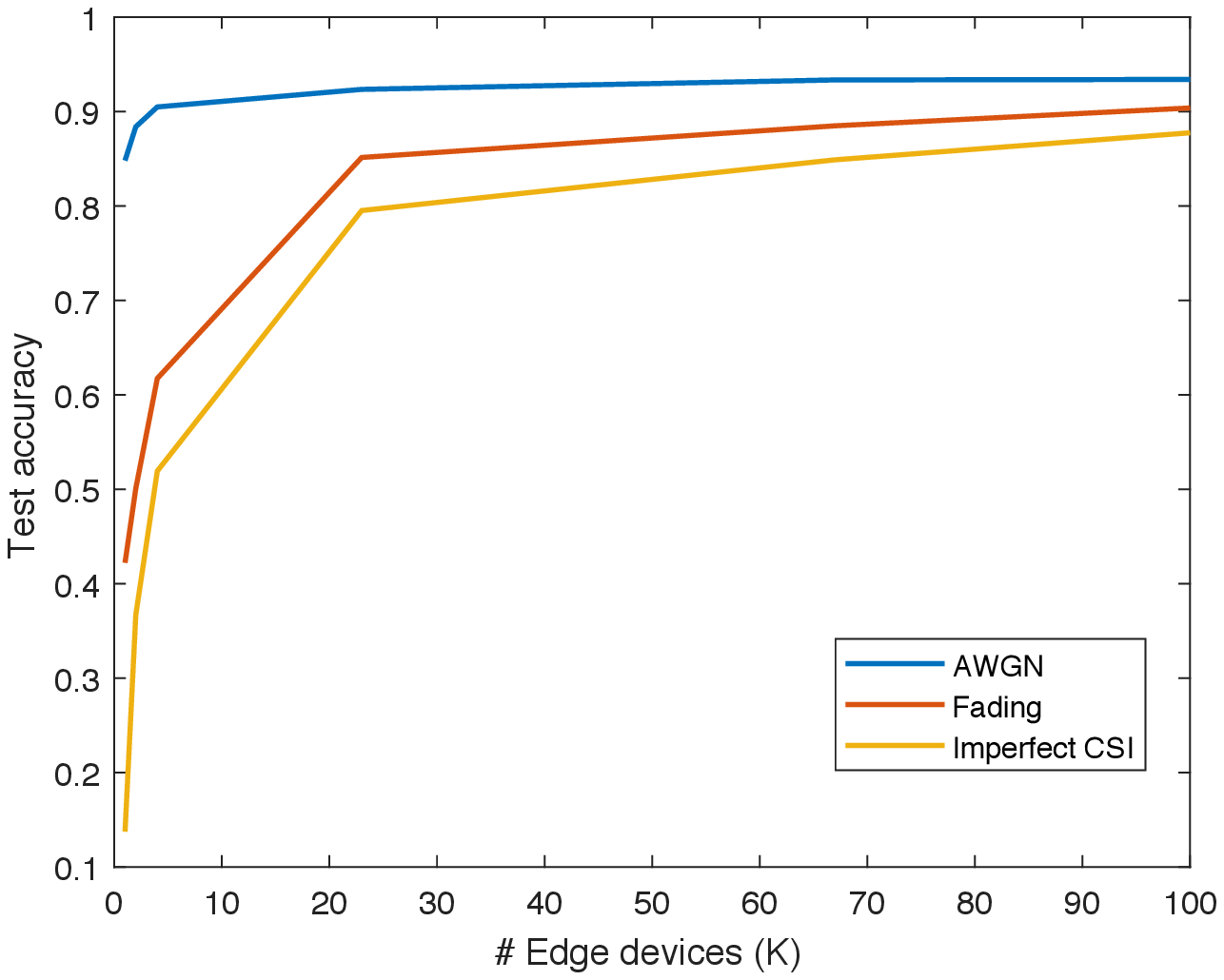}}
\subfigure[Dataset: CIFAR10; Classifier: ResNet18]{\includegraphics[width=0.42\textwidth]{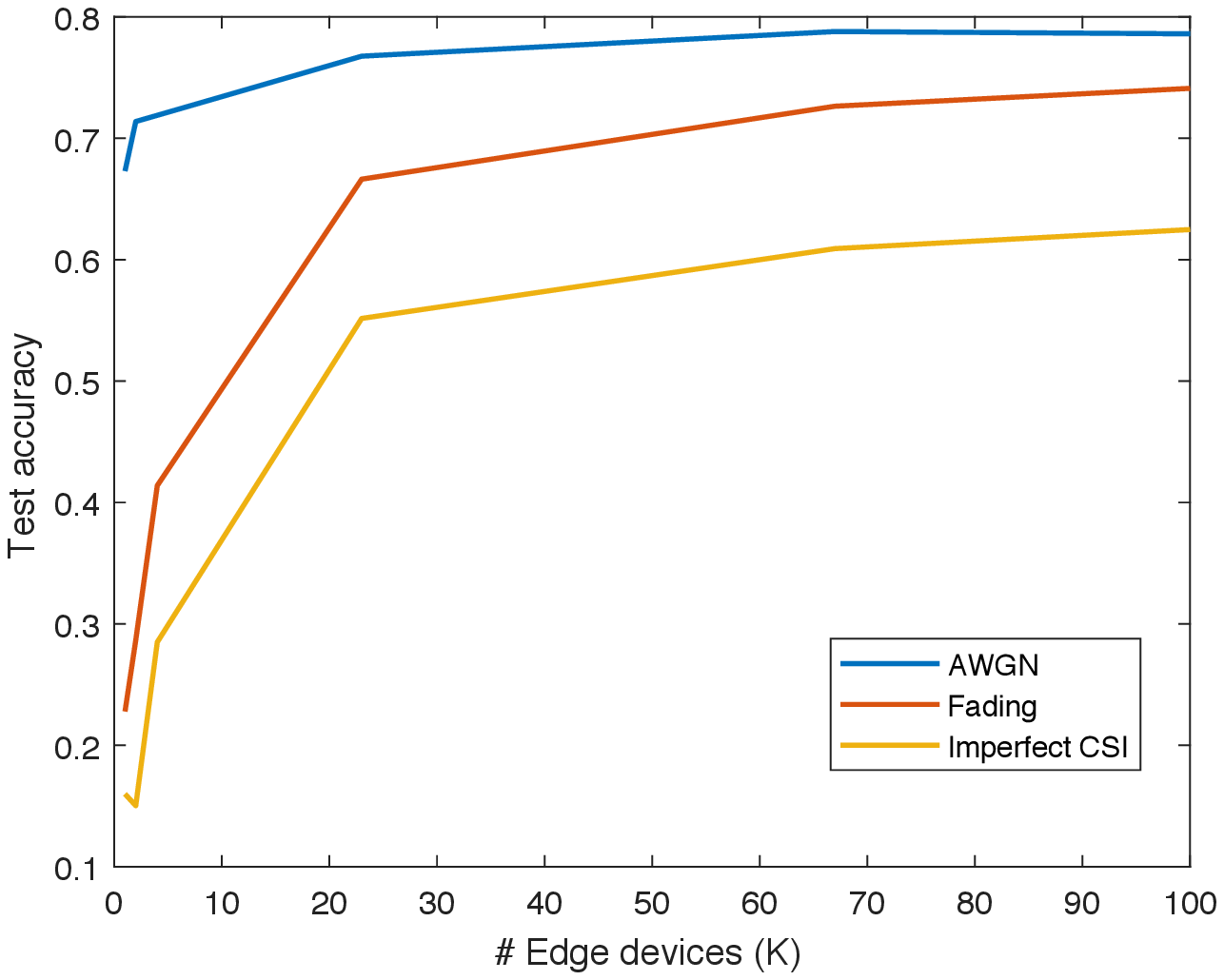}}
\vspace{3mm}
\caption{Effect of  device population on the convergence.
 }
\label{Fig:effect_user_number}
\end{figure}

The effect of the device population on the convergence behaviour is illustrated in Fig.~\ref{Fig:effect_user_number}, where we set the number of communication-rounds as $150$ for the MNIST dataset and $200$ for the CIFAR10 dataset, and plot the curves of test accuracy w.r.t. the total number of edge-devices $K$ for the three considered scenarios. It is observed that, in all scenarios, the test accuracy grows as $K$ increases. This is because  a larger $K$ suppresses the noise variance inherent in stochastic gradients as well as  the negative effects due to wireless hostilities. This phenomenon is coined as \emph{majority-vote gain}. In particular, the majority-vote gain is observed to be the most prominent in the gentle wireless condition (i.e., AWGN), and weakened in the hostile one (i.e., fading with imperfect CSI).
This behaviour is aligned with analytical results in Theorems \ref{theo:AWGN}-\ref{theo:imperfect_CSI}, which showed that the negative effects introduced by different wireless hostilities vanish, at different rates, with the growth of the device population.

\setlength{\textfloatsep}{4pt}

\setlength{\textfloatsep}{4pt}


\vspace{-1mm}
\subsection{Performance Comparison: OFDMA, BAA  and OBDA}
\vspace{-1mm}
 
\begin{figure}[tt]
\centering
\subfigure[Test accuracy ]{\label{Fig:test_acc}\includegraphics[width=0.42\textwidth]{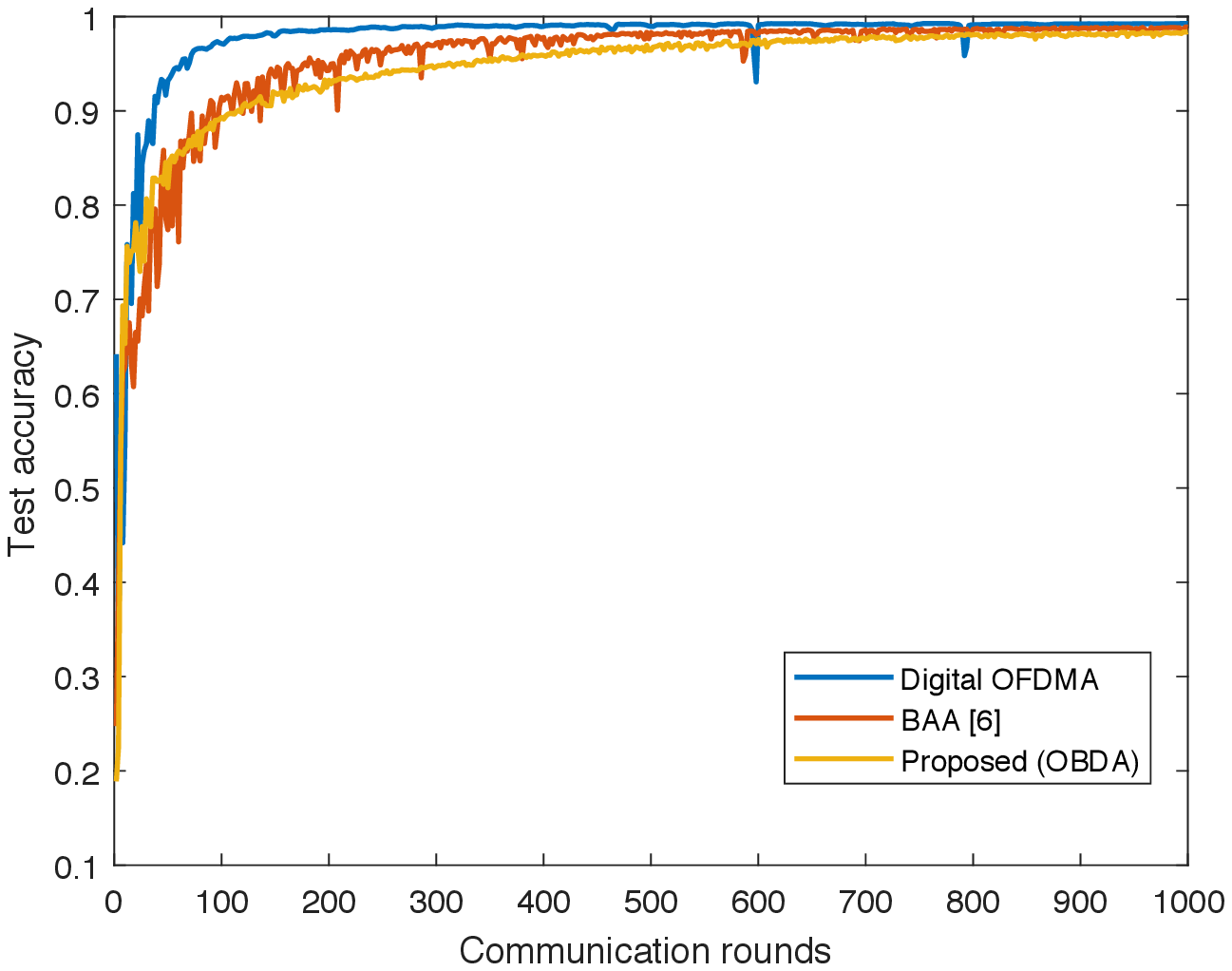}}
\subfigure[Communication latency]{\label{Fig:latency}\includegraphics[width=0.42\textwidth]{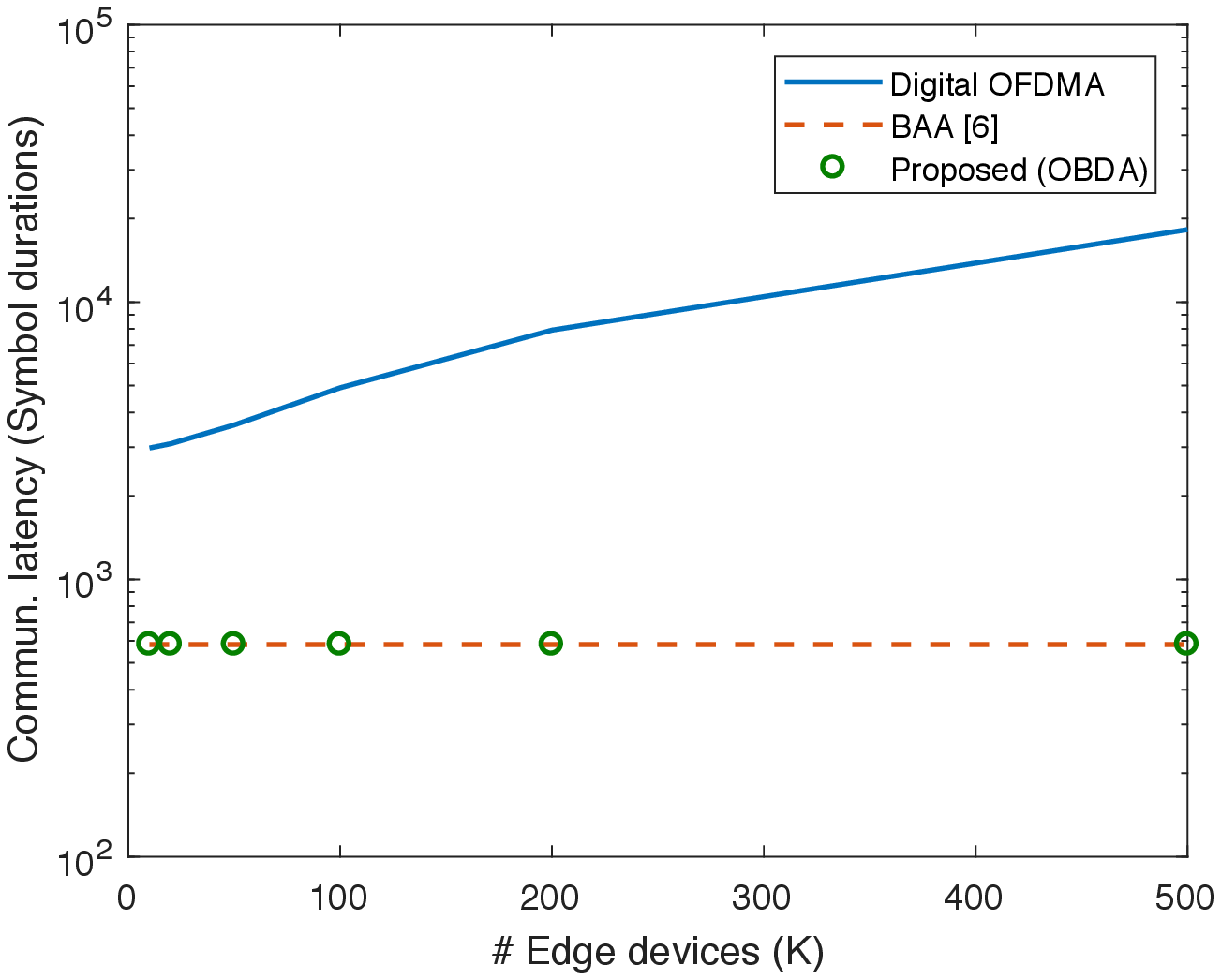}}
\vspace{3mm}
\caption{Performance comparison among digital OFDMA, BAA and OBDA.
 }
\label{Fig:comparison}
\end{figure}

A performance comparison between digital transmission using OFDMA, BAA developed in  \cite{zhu2018low}, and the proposed OBDA  is presented in Fig. \ref{Fig:comparison}. In this figure, the test accuracy and communication latency  are plotted for these three  schemes over a fading MAC with perfect CSI. For the digital OFDMA, sub-channels are evenly allocated to the edge devices; 
 gradient-update parameters are quantized into  bit sequences with 16-bit per coefficient; and  adaptive MQAM modulation is used to maximize the spectrum efficiency under a target bit error rate of $10^{-3}$. 
 It can be observed from Fig. \ref{Fig:test_acc} that the convergence speed  of digital OFDMA, BAA and OBDA are in descending order while all three schemes achieve comparable accuracies after sufficient number of communication rounds. The reason behind the faster convergence  of  digital OFDMA w.r.t. BAA is that the direct exposure of the analog modulated signals in BAA to the channel hostilities results in a boosted expected gradient norm as mentioned in Remarks 1-3. The performance gap between the BAA and OBDA in terms of  convergence speed arises from the quantization loss introduced by the latter. 
However, we observe that the gap between the two is small, which shows that over-the-air gradient aggregation can be employed with devices employing simple 4-QAM modulation without significant performance loss w.r.t. analog aggregation. Moreover, Fig. \ref{Fig:latency} shows that, without compromising the learning accuracies, the \emph{per-round communication latencies} for both OBDA and BAA are independent of the number of devices, while that of the digital OFDMA scales up as the device 
population grows.  

\setlength{\textfloatsep}{4pt}

\section{Concluding Remarks}\label{Conclusion}
In the context of FEEL, we have proposed a novel digital over-the-air gradient aggregation scheme, called 
 OBDA, for communication-efficient distributed learning across wireless devices. To understand its performance, a comprehensive convergence analysis framework for OBDA subject to wireless channel hostilities is developed. This work represents the first attempt to develop digital AirComp, which is more practical,  compared to its analog counterpart, in terms of the compatibility with the modern digital communication infrastructure. For future work, we will consider the generalization of the current work to multi-cell FEEL, where the effect of inter-cell interference should also be taken into account. 
 As another interesting direction, the proposed design assuming a single learning task can be extended to the multi-task learning scenario, where an additional task scheduler at the server needs to be designed in an effort to reducing the frequency of the gradient updates, further accelerating the learning process.

\appendix
\subsection{Proof of Theorem \ref{theo:AWGN} 
} \label{app:theo:AWGN}
The proof is conducted following the widely-adopted strategy of relating the norm of the gradient to the expected improvement made in a single algorithmic step, and comparing this with the total possible improvement under Assumption 1. A key technical challenge we overcome is in showing how to directly deal with a \emph{biased gradient estimate} ${\bf v}$ of the full-batch gradient ${\bf g}$ as specified in the sequel. Note that the current subsection also serves as another purpose of presenting the general framework for convergence analysis of OBDA, which also applies to the later extension to the more complicated scenarios. 

To start with, we first bound the improvement of the objective during a single step of the algorithm for one instantiation of the data-stochasticity induced noise based on Assumption 2. To this end, substituting the step in \eqref{eq:model_update_onebit} to \eqref{eq:smoothness} and decomposing the improvement to expose the (data-and-channel) stochasticity-induced error we have:
\begin{align}
& F^{(n+1)} - F^{(n)} \notag \\
\leq & (\bg^{(n)})^T (\bw^{(n+1)} - \bw^{(n)}) + \frac{1}{2} \sum_{i=1}^q L_i \l(w^{(n+1)}_i - w^{(n)}_i\r)^2, \notag\\
 = &-\eta (\bg^{(n)})^T {\sf sign (\tilde \bg^{(n)})} + \eta^2 \sum_{i=1}^q \frac{L_i}{2}, \notag\\
 = & -\eta \|\bg^{(n)}\|_1 + \frac{\eta^2}{2} \|\bL\|_1  + \notag \\ 
   &\qquad \qquad  \underbrace{2\eta \sum_{i=1}^q |g^{(n)}_i| {\mathbb I} [{\sf sign}(\tilde g^{(n)}_i) \neq {\sf sign}(g^{(n)}_i)]}_{\sf Stochasticity-induced\;\; error}, 
\end{align}
where $\tilde g^{(n)}_i$  and $g^{(n)}_i$ denote the $i$-th element of $\tilde \bg^{(n)}$ and $\bg^{(n)}$, respectively. 

Next we find the expected improvement at time $n+1$ conditioned on the previous iterate.
\begin{multline}\label{eq:conditioned_expected_improvement}
\mathbb{E} [F^{(n+1)} - F^{(n)} | \bw^{(n)}] \leq -\eta \|\bg^{(n)}\|_1 + \frac{\eta^2}{2} \|\bL\|_1 + \\ \underbrace{2\eta \sum_{i=1}^q |g^{(n)}_i| {\mathbb P} [{\sf sign}(\tilde g^{(n)}_i) \neq {\sf sign}(g^{(n)}_i)]}_{\sf Stochasticity-induced\;\; error},
\end{multline}
where $\bg^{(n)}$ is a constant vector due to the conditioning.
It can be noted from \eqref{eq:conditioned_expected_improvement} that the expected improvement on the objective crucially depends on the \emph{(decoding) bit error probability}, i.e.,
\begin{align} \label{def:err_prob}
P^{\sf err}_i =  {\mathbb P} [{\sf sign}(\tilde g^{(n)}_i) \neq {\sf sign}(g^{(n)}_i)],
\end{align}
which is intuitively determined by the level of the noise introduced by the data-stochasticity and the wireless channel. To formalize the intuition, we have the following lemma. 
\begin{lemma}\label{lemma:error_prob_AWGN} 
\emph{The bit error probability in the AWGN channel is bounded by}
\begin{align}
P^{\sf err}_i \leq \frac{1}{\sqrt{K}S_i} + \frac{\sigma_z}{KS_i\sqrt{\rho_0}} + \frac{\sigma_z}{2K \sqrt{\rho_0}},
\end{align}
\emph{where we have defined $S_i = \sqrt{n_b} \frac{|g_i^{(n)}|}{\sigma_i}$ as the \emph{gradient-signal-to-data-noise ratio}. The coefficient $\sqrt{n_b}$ is because each of the local gradient estimate $\tilde g_{k,i}$ is computed over a mini-batch of size $n_b$, thus the resultant gradient variance reduces from $\sigma_i^2$ to $\frac{\sigma_i^2}{n_b}$ according to Assumption 3 and Equation \eqref{eq:local_update}.}
\end{lemma}
\proof  
See Appendix \ref{app:lemma:error_prob_AWGN}
\endproof
 
Then, substituting Lemma \ref{lemma:error_prob_AWGN} into \eqref{eq:conditioned_expected_improvement} we have
\begin{align}\label{eq:conditioned_expected_improvement_2}
& \mathbb{E} [F^{(n+1)} \!-\! F^{(n)} | \bw^{(n)}] \notag \\
 \leq & -\eta \|\bg^{(n)}\|_1 \!+\! \frac{\eta^2}{2} \|\bL\|_1 +  \notag  \\
  &  \qquad \;\; 2\frac{\eta}{\sqrt{n_b}}\l(\frac{1}{\sqrt{K}} \!+\! \frac{\sigma_z}{K \sqrt{\rho_0}}\r) \|\boldsymbol \sigma\|_1 +  \frac{\eta \sigma_z}{K \sqrt{\rho_0}}\|\bg^{(n)}\|_1,   \notag \\
 = & \eta \l(\frac{\sigma_z}{K \sqrt{\rho_0}} \!-\! 1\r)\|\bg^{(n)}\|_1 \!+\! \frac{\eta^2}{2} \|\bL\|_1 + \notag \\  & \qquad \qquad \qquad \qquad   2\frac{\eta}{\sqrt{n_b}}\l(\frac{1}{\sqrt{K}} \!+\! \frac{\sigma_z}{K \sqrt{\rho_0}}\r) \|\boldsymbol \sigma\|_1. \!\!\!
\end{align}

{
Further plug in the learning rate and mini-batch settings $\eta = \frac{1}{\sqrt{\|{\bf L}\|_1 n_b}}$ and $n_b = \frac{1}{\gamma}N$, we have
\begin{multline}
\mathbb{E} [F^{(n+1)} \!-\! F^{(n)} | \bw^{(n)}] \!\leq\! \frac{1}{\sqrt{\|{\bf L}\|_1 N}} \l(\frac{\sigma_z}{K \sqrt{\rho_0}} \!-\! 1\r)\|\bg^{(n)}\|_1 + \\
 \frac{\gamma}{2N}  +  \frac{2\gamma}{\sqrt{\|{\bf L}\|_1} N} \l(\frac{1}{\sqrt{K}} \!+\! \frac{\sigma_z}{K \sqrt{\rho_0}} \r) \|\boldsymbol \sigma\|_1. \nn
\end{multline}

Now we take the expectation over $\bw^{(n)}$ to average out the randomness in the optimization trajectory and perform a telescoping sum over the iterates:
\begin{align} \label{eq:bounded_improvement}
&F^{(0)} - F^* \nn \\
\geq& F^{(0)} - {\mathbb E}[F^{(n)}]  = \mathbb{E}\l[\sum_{n=0}^{N-1} F^{(n)} - F^{(n+1)} \r], \nn\\
 \geq& \mathbb{E} \l[ \sum_{n=0}^{N-1}  \frac{1}{\sqrt{\|{\bf L}\|_1 N}} \l(1 - \frac{\sigma_z}{K \sqrt{\rho_0}} \r)\|\bg^{(n)}\|_1 - \right. \notag \\
  &\qquad \left. \frac{\gamma}{2\sqrt{\|{\bf L}\|_1} N} \l( (\frac{4}{\sqrt{K}} + \frac{4\sigma_z}{K \sqrt{\rho_0}}) \|\boldsymbol \sigma\|_1 + \sqrt{\|{\bf L}\|_1} \r) \r], \nn \\
 =& \sqrt{\frac{N}{\|\bL\|_1}} \l(1 - \frac{\sigma_z}{K \sqrt{\rho_0}} \r) \mathbb{E}\l[\frac{1}{N} \sum_{n=0}^{N-1}  \|\bg^{(n)}\|_1 \r]    - \notag \\
  & \qquad \frac{\gamma}{2\sqrt{\|{\bf L}\|_1}} \l( (\frac{4}{\sqrt{K}} + \frac{4\sigma_z}{K \sqrt{\rho_0}}) \|\boldsymbol \sigma\|_1 + \sqrt{\|{\bf L}\|_1} \r).
\end{align}

In the end, the desired result in Theorem \ref{theo:AWGN} can be easily obtained by rearranging the terms in \eqref{eq:bounded_improvement}, which completes the proof. 
}

\subsection{Proof of Lemma  \ref{lemma:error_prob_AWGN} 
}\label{app:lemma:error_prob_AWGN} 
The key idea of the proof is to establish an equivalent mathematical event of ${\sf sign}(\tilde g^{(n)}_i) = {\sf sign}(g^{(n)}_i)$ described by well-defined random variables with known distributions. To this end, let $X_i$ denote the number of edge devices with correct sign bit at the $i$-th element of the gradient vector, namely, that with ${\sf sign}(g^{(n)}_{k,i}) = {\sf sign}(g^{(n)}_i)$, and 
$\tilde X_i = X_i + \tilde z_i$ denote the noisy version of $X_i$ corrupted by the effective channel noise $\tilde z_i = \frac{z_i}{2\sqrt{\rho_0}} \sim {\cal N}(0, \frac{\sigma_z^2}{4\rho_0})$. Note that  $X_i$ is the sum of $K$ independent Bernoulli trials, and thus binomial with success probability and failure probability denoted by 
\begin{align}
p_i &= \mathbb{P} [{\sf sign}( g^{(n)}_{k,i}) = {\sf sign}(g^{(n)}_i)],  \\
q_i &= \mathbb{P} [{\sf sign}( g^{(n)}_{k,i}) \neq {\sf sign}(g^{(n)}_i)],
\end{align}
respectively. Then we derive the mean and variance of $\tilde X_i$ as follows:
\begin{align}
{\mathbb E} [\tilde X_i] &= K p_i = K \l(\epsilon_i + \frac{1}{2} \r), \label{mean_t_x} \\
 {\sf Var}(\tilde X_i) &= K p_i q_i + \frac{\sigma_z^2}{4\rho_0} = K \l(\frac{1}{4} -\epsilon_i^2 \r) + \frac{\sigma_z^2}{4\rho_0}, \label{var_t_x}
\end{align}
 where we have defined $\epsilon_i = p_i - 1/2 = 1/2 - q_i$ for ease of subsequent derivation.

According to \eqref{eq:aggregated_gradient_AWGN}, to ensure that 
\begin{align}
{\sf sign}(\tilde g^{(n)}_i) = {\sf sign}\l(\sum_{k =1}^K \sqrt{\rho_0} \tilde g^{(n)}_{k,i} + z_i\r) = {\sf sign}(g^{(n)}_i),
\end{align}
$\tilde X_i$ must be larger than $\frac{K}{2}$. Therefore we have
\begin{align}\label{eq:equivalent_def_error_prob}
P^{\sf err}_i = {\mathbb P}\l({\tilde X_i} \leq \frac{K}{2}\r) = {\mathbb P} \l[{\mathbb E}[\tilde X_i] - \tilde X_i \geq {\mathbb E}[\tilde X_i] - \frac{K}{2} \r].
\end{align} 
Applying the known Cantellis' inequality, ${\mathbb P} (X - {\mathbb E}[X] \geq \lambda) \leq \frac{{\sf var}(X)}{{\sf var}(X) + \lambda^2}$, $\lambda > 0$, on \eqref{eq:equivalent_def_error_prob} yields
\begin{align}
P^{\sf err}_i \leq \frac{1}{ 1 + \frac{({\mathbb E}[\tilde X_i] - K/2)^2}{{\var(\tilde X_i)}}} \leq \frac{1}{2} \sqrt{\frac{\var(\tilde X_i)}{\mathbb E[\tilde X_i] - \frac{K}{2}}},
\end{align}
where the second inequality is due to the fact that $1 + a^2 \geq 2a$. 
Next, we plug in the statistics of $\tilde X$ in \eqref{mean_t_x} and \eqref{var_t_x}, to obtain
\begin{align}\label{eq:bound_p_err}
P^{\sf err}_i &\leq \frac{1}{2} \sqrt{\frac{1}{K} \l(\frac{1}{4\epsilon_i^2} - 1\r) + \frac{\sigma_z^2}{K^2 \epsilon_i^2 4 \rho_0} }, \nn \\
& \leq \frac{1}{2} \sqrt{\frac{1}{K} \l(\frac{1}{4\epsilon_i^2} - 1\r)} + \frac{1}{2} \frac{\sigma_z}{\epsilon_i K \sqrt{4\rho_0}},
\end{align}
where the second inequality is due to the fact that $\sqrt{a + b} \leq \sqrt{a} + \sqrt{b}$.

To proceed with, we need a bound on $\epsilon_i = 1/2 -q_i$ that can relate it to the gradient-signal-to-data-noise ratio $S_i$ defined in Lemma \ref{lemma:error_prob_AWGN}.  The bound can be attained from the following bound on $q_i$, the failure probability for the sign bit of a single device, under the unimodal symmetric gradient noise assumption stated in Assumption 4.
\begin{lemma}\label{lemma:bound_fail_prob} 
\emph{Under the assumption of unimodal symmetric gradient noise in Assumption 4, the failure probability for the sign bit of a single device can be bounded by }
\begin{align} 
q_i &= \mathbb{P}\l[ {\sf sign}(\tilde g^{(n)}_{k,i}) \neq {\sf sign}(g^{(n)}_i)\r], \nn \\
&\leq
\l\{\begin{aligned} 
&\frac{2}{9} \frac{1}{S_i^2}, && \text{if} \; S_i > \frac{2}{\sqrt{3}} \\
&\frac{1}{2} - \frac{S_i}{2\sqrt{3}}, && \text{otherwise}. 
\end{aligned}
\r. \qquad \forall i.
\end{align}
\end{lemma}
\proof
The result follows from the known Gauss' inequality recalled below. For a unimodal symmetric random variable $X$ with mean $\mu$ and variance $\sigma^2$, the below inequality holds:
\begin{align}\label{eq:Gauss_ineq}  \qquad
\mathbb{P}[|X - \mu| > x]
\leq
\l\{\begin{aligned} 
&\frac{4}{9} \frac{\sigma^2}{x^2}, && \text{if} \; \frac{x}{\sigma} > \frac{2}{\sqrt{3}} \\
&1 - \frac{x}{\sqrt{3}\sigma}, && \text{otherwise}. 
\end{aligned}
\r.
\end{align}
Without loss of generality, assume that $g^{(n)}_{i}$ is negative. Then applying the Gauss' inequality, the bound on the failure probability for the sign bit can be derived as follows: 
\begin{align}\label{eq:bound_fail_prob}
q_i &= \mathbb{P}\l[ {\sf sign}(\tilde g^{(n)}_{k,i}) \neq {\sf sign}(g^{(n)}_i)\r]  =  \mathbb{P}\l[\tilde g^{(n)}_{k,i} - g^{(n)}_i \geq |g^{(n)}_i| \r], \nn \\
& = \frac{1}{2} \mathbb{P}\l[ |\tilde g^{(n)}_{k,i} - g^{(n)}_i| \geq |g^{(n)}_i| \r], \nn \\
& \leq \l\{\begin{aligned} 
&\frac{2}{9} \frac{\sigma_i^2}{n_b |g^{(n)}_i|^2}, && \text{if} \; \frac{|g^{(n)}_i|}{\sigma_i/\sqrt{n_b}} > \frac{2}{\sqrt{3}} \\
&\frac{1}{2} - \frac{ |g^{(n)}_i|}{2\sqrt{3}\sigma_i/\sqrt{n_b}}, && \text{otherwise},
\end{aligned}
\r.
\end{align}
where the term $\sqrt{n_b}$ is due to that each of the local gradient estimate $\tilde g_{k,i}$ is computed over a mini-batch of size $n_b$. Finally, the desired result is obtained by noting $S_i = \sqrt{n_b} \frac{|g_i^{(n)}|}{\sigma_i}$.
\endproof
Next, combining the equation $\epsilon_i = 1/2 -q_i$ and Lemma \ref{lemma:bound_fail_prob}, we can obtain:
\vspace{-3mm}
\begin{align}\label{eq:bound_on_epsilon}
\vspace{-2mm}
\frac{1}{4\epsilon_i^2} - 1 \leq \frac{4}{S_i^2},
\end{align}
whose proof involves only simple algebraic manipulations, e.g., checking the monotonicity of the piece-wise function on the right hand side of \eqref{eq:bound_fail_prob}, and is skipped here for brevity.  Note that \eqref{eq:bound_on_epsilon} further implies that 
\begin{align}\label{eq:bound_on_epsilon_2}
\frac{1}{\epsilon_i} \leq 2 \sqrt{\frac{4}{S_i^2} + 1} \leq \frac{4}{S_i} + 2,
\end{align}
which follows from the fact that $\sqrt{a + b} \leq \sqrt{a} + \sqrt{b}$.
Finally, by substituting \eqref{eq:bound_on_epsilon} and \eqref{eq:bound_on_epsilon_2} into \eqref{eq:bound_p_err}, the error probability of received sign vector can be further bounded by
\begin{align}
P^{\sf err}_i \leq \frac{1}{\sqrt{K} S_i} + \l( \frac{2}{S_i} + 1\r) \frac{\sigma_z}{K\sqrt{4\rho_0}}.
\end{align}
This gives the desired result by simply rearranging the terms. 

\subsection{Proof of Theorem \ref{theo:fading} 
} \label{app:theo:fading}
The convergence analysis for the fading channel case  also follows the general strategy of relating the norm of the gradient to the expected improvement in objective as presented in Appendix \ref{app:theo:AWGN}. Specifically, the expression for calculating the single-step expected improvement in \eqref{eq:conditioned_expected_improvement} still applies for the fading channel case, while a new expression for bit error probability $P^{\sf err}_i$ defined in \eqref{def:err_prob} need be derived to account for the additional randomness introduced by channel fading. This is the key challenge tackled in the following proof. 

Due to the adoption of the truncated channel inversion power control in \eqref{truncated_channel_inv}, the random channel fading makes the devices sending the $i$-th gradient element a random set denoted by ${\cal K}_i$, with its size denoted by $K_i$. We can rewrite the $i$-th aggregated gradient element at the channel output in \eqref{channel_model} as follows:
\begin{align}\label{eq:aggregated_gradient_fading}
\tilde g_i^{(n)} = \sum_{k \in {\cal K}_i} \sqrt{\rho_0} \tilde g_{k,i}^{(n)} + z_i^{(n)}, 
\end{align}

Conditioned on the size of the transmitting set $K_i$, the error probability of received sign bit is first derived. Based on this, the unconditional error probability will be derived next. 
\begin{lemma}\label{lemma:conditioned_err_prob} \emph{Consider the fading channel with truncated channel inversion power control with perfect CSI. The bit error probability conditioned on the size of the transmitting set $K_i$ is given by}
\begin{align}
P^{\sf err}_i (K_i) &= {\mathbb P} [{\sf sign}(\tilde g^{(n)}_i) \neq {\sf sign}(g^{(n)}_i) | {K}_i],  \\ 
&\leq \l\{\begin{aligned} 
&  \frac{1}{\sqrt{K_i}S_i} + \frac{1}{K_i} \frac{\sigma_z}{\sqrt{\rho_0}} \l( \frac{1}{S_i} + \frac{1}{2}\r), && K_i \geq 1 \\
&  \frac{1}{2}, && K_i = 0. 
\end{aligned}
\r. \nn
\end{align}
\end{lemma}
\proof
For the case with non-empty transmitting set, i.e., $K_i \geq 1$, the result directly follows Lemma \ref{lemma:error_prob_AWGN} by noting the similarity between the channel model in \eqref{eq:aggregated_gradient_fading} and that in \eqref{eq:aggregated_gradient_AWGN}. While for the case  $K_i = 0$, no device is transmitting, and thus only channel noise is received at the edge server, thereby the  decoding process reduces to a random guess, with an error probability of $\frac{1}{2}$.  
\endproof

We note that the size of the transmitting set, $K_i$ follows a binomial distribution $K_i \sim B(K, \alpha)$ with $\alpha$ denoting the non-truncation probability derived in \eqref{eq:truncation_ratio}. This immediately leads to the following results:
\begin{align}\label{eq:prob_of_K_i}
\mathbb{P} (K_i = 0) & = (1-\alpha)^K, \\ 
\mathbb{P} (K_i \geq 1) 
& = \sum_{k=1}^K {K \choose k} \alpha^k (1-\alpha)^{K-k}.
\end{align}

By the law of total probability, the unconditioned error probability $P^{\sf err}_i$ can be computed via 
\begin{align}\label{eq:unconditioned_error_prob}
P^{\sf err}_i = P^{\sf err}_i(K_i = 0)   \mathbb{P} (K_i = 0) + P^{\sf err}_i(K_i \geq 1)  \mathbb{P} (K_i \geq 1).
\end{align}

Then, by substituting  Lemma \ref{lemma:conditioned_err_prob} and the results in \eqref{eq:prob_of_K_i} into \eqref{eq:unconditioned_error_prob}, we can compute a bound on the unconditional error probability as follows: 
\begin{multline}\label{eq:unconditioned_error_prob_bound}
P^{\sf err}_i \leq \frac{1}{2} (1-\alpha)^K +  \sum_{k=1}^K {K \choose k} \alpha^k (1-\alpha)^{K-k} \times \\
\l[ \frac{1}{\sqrt{k}S_i} + \frac{1}{k} \frac{\sigma_z}{\sqrt{\rho_0}} \l( \frac{1}{S_i} + \frac{1}{2}\r) \r]. 
\end{multline}
To simplify \eqref{eq:unconditioned_error_prob_bound}, we find it useful to establish the following two important inequalities.
\begin{lemma}\label{lemma:two_inequalities}
\emph{The following two inequalities hold:}
\begin{align}
f(K,\alpha) & \equiv	 \sum_{k=1}^K\frac{1}{k}{K \choose k} \alpha^k (1-\alpha)^{K-k} \leq \frac{2}{K\alpha} \label{eq:bounded_f}, \\
g(K,\alpha) & \equiv	 \sum_{k=1}^K\frac{1}{\sqrt{k}}{K \choose k} \alpha^k (1-\alpha)^{K-k} \leq \frac{\sqrt{6}}{\sqrt{K\alpha}}.
\end{align}
\end{lemma}
\proof
We start with the first inequality. Function $f(K,\alpha)$ can be  rewritten as follows: 
\begin{align}
& f(K,\alpha) \nn\\
= & K\alpha  \sum_{k=1}^K\frac{1}{k^2}{K-1 \choose k-1} \alpha^{k-1} (1-\alpha)^{K-k},  \nn\\
 = & K\alpha \sum_{k=0}^{K-1}\frac{1}{(k+1)^2}{K-1 \choose k} \alpha^{k} (1-\alpha)^{K-k-1}, \nn\\
 = & K\alpha \sum_{k=0}^{K-1}\frac{k+2}{k+1}\frac{1}{(k+1)(k+2)}{K-1 \choose k} \alpha^{k} (1-\alpha)^{K-k-1}. \nn
\end{align}
Note that since $\frac{k+2}{k+1} \leq  2$, the function $f(K,\alpha)$ is  bounded by \eqref{eq:bounded_f_2} as shown on the top of next page. 
\begin{figure*}
\begin{align} \label{eq:bounded_f_2}
f(K,\alpha) &\leq  \frac{2 K \alpha}{K(K+1)\alpha^2}  \sum_{k=0}^{K-1} {K+1 \choose k+2} \alpha^{k+2} (1-\alpha)^{K-k-1},            \\
& = \frac{2}{(K+1)\alpha}  \l[ \overbrace{\underbrace{\sum_{k=-2}^{K-1} {K+1 \choose k+2} \alpha^{k+2} (1-\alpha)^{K-k-1}}_{=1} 
- (1-\alpha)^{K+1} - (K+1)\alpha(1-\alpha)^K}^{<1}\r]  \leq \frac{2}{(K+1)\alpha} \leq \frac{2}{K\alpha}. \nn
\end{align}
\hrule
\vspace{-3mm}
\end{figure*}

Next, we move to the proof of the second inequality. Similarly, we have
\begin{align}\label{app:bound_g}
g(K,\alpha) &\!=\! K\alpha  \sum_{k=1}^K\frac{1}{k\sqrt{k}}{K-1 \choose k-1} \alpha^{k-1} (1-\alpha)^{K-k},  \\
& \!=\! K\alpha \sum_{k=0}^{K-1}\l(\frac{1}{k+1}\r)^{\frac{3}{2}}{K-1 \choose k} \alpha^{k} (1-\alpha)^{K-k-1}. \nn
\end{align}
We remark that the undesired square root operation on $\frac{1}{k+1}$ makes it harder to derive a bound on $g(K,\alpha)$ as the tricks used for deriving \eqref{eq:bounded_f} cannot be applied here. Nevertheless the challenge can be overcome by rewritting \eqref{app:bound_g} as 
\begin{align}\label{app:alternative_ex_g}
g(K,\alpha) = K\alpha {\mathbb E} \l[\l(\frac{1}{k+1}\r)^{\frac{3}{2}}\r],
\end{align}
where the expectation is taken over a random variable $k$ that follows the binomial distribution $B(K-1, \alpha)$.
Then by Jensen's inequality, we have 
\begin{align}\label{app:Jensen's inequality}
{\mathbb E} \l[ \l(\frac{1}{k+1}\r)^{\frac{3}{2}} \r] \leq \sqrt{{\mathbb E} \l(\frac{1}{k+1}\r)^3}.
\end{align}
 This suggests that we can obtain a bound on $g(K, \alpha)$ by bounding ${\mathbb E} \l[ \l(\frac{1}{k+1}\r)^3 \r]$, which is derived below. We have
\begin{multline}
{\mathbb E} \l[ \l(\frac{1}{k+1}\r)^3 \r] = K\alpha \sum_{k=0}^{K-1}\frac{k+3}{k+1}\frac{k+2}{k+1} \times\\
\frac{1}{(k+1)(k+2)(k+3)}{K-1 \choose k} \alpha^{k} (1-\alpha)^{K-k-1}.
\end{multline}
Note that since $\frac{k+3}{k+1}\frac{k+2}{k+1} \leq  6$, the expectation can be further bounded by \eqref{app:bound_E_k3} as shown on the top of next page.
\begin{figure*}
\begin{align}\label{app:bound_E_k3}
{\mathbb E} \l[ \l(\frac{1}{k+1}\r)^3 \r] & \! \leq \! \frac{6}{K(K+1)(K+2)\alpha^3}  \sum_{k=0}^{K-1} {K+2 \choose k+3} \alpha^{k+3} (1-\alpha)^{K-k-1}            \\
& \!=\! \frac{6 \l[ 1 - (1-\alpha)^{K+2} - (K+2)\alpha(1-\alpha)^{K+1} - \frac{(K+2)(K+1)}{2}\alpha^2(1-\alpha)^K \r]}{K(K+1)(K+2)\alpha^3}   \! \leq \! \frac{6}{K(K+1)(K+2)\alpha^3} \! \leq \! \frac{6}{K^3\alpha^3}. \nn
\end{align}
\hrule
\vspace{-3mm}
\end{figure*}
Finally, plugging \eqref{app:bound_E_k3} and \eqref{app:Jensen's inequality} into \eqref{app:alternative_ex_g} we can attain the desired second inequality.
\endproof

Then applying Lemma \ref{lemma:two_inequalities} on \eqref{eq:unconditioned_error_prob_bound} gives the unconditional bit error probability as follows.
\begin{lemma}\label{lemma:unconditioned_err_prob} \emph{Consider the fading channel with truncated channel inversion power control with perfect CSI. The unconditional bit error probability is bounded by}
\begin{align}
P^{\sf err}_i \leq \frac{1}{2} (1-\alpha)^K +   \frac{\sqrt{6}}{\sqrt{\alpha K}S_i} + \frac{2}{\alpha K} \frac{\sigma_z}{\sqrt{\rho_0}} \l( \frac{1}{S_i} + \frac{1}{2}\r) .
\end{align}
\end{lemma}
With Lemma \ref{lemma:unconditioned_err_prob} at hand, the desired result in Theorem \ref{theo:fading} can be easily derived by substituting Lemma \ref{lemma:unconditioned_err_prob} into \eqref{eq:conditioned_expected_improvement} and following the same machinery presented in \eqref{eq:conditioned_expected_improvement_2}-\eqref{eq:bounded_improvement}.

\vspace{-4mm}
\subsection{Proof of Theorem \ref{theo:imperfect_CSI} 
} \label{app:theo:imperfect_CSI}
\vspace{-1mm}
Again, the same analysis framework presented in Appendix \ref{app:theo:AWGN} applies, while the remaining work is to derive a new bit error probability $P^{\sf err}_i$ (defined in \eqref{def:err_prob}) to account for the additional error introduced by the imperfect CSI as presented in the following. 

First, we define ${\cal K}_i = \{ k \mid |\hat h_k|^2 \geq g_{\sf th} \}$ as the set of devices transmitting the $i$-th gradient element, whose estimated channel gain is larger than the cutoff threshold.  Then, according to the imperfect CSI model in \eqref{eq:channel_estimation_error}, we can rewrite the $i$-th aggregated gradient element at the channel output in \eqref{channel_model} as follows:
\begin{align}\label{eq:aggregated_gradient_fading_im_CSI}
\tilde g_i^{(n)} = \sum_{k \in {\cal K}_i} h_k \frac{\sqrt{\rho_0}}{\hat h_k} \tilde g_{k,i}^{(n)} + z_i = \sum_{k \in {\cal K}_i}  \frac{\sqrt{\rho_0}}{1 + \frac{\Delta}{h_k}}  \tilde g_{k,i}^{(n)} + z_i.
\end{align}

Next, according to the assumption $|\Delta| \leq \Delta_{\max}\ll \sqrt{g_{\sf th}}$, and the fact that $|\hat h_k| \geq \sqrt{g_{\sf th}}$, for $k \in {\cal K}_i$, we can show that 
\begin{align}\label{app:del_over_hk}
\frac{|\Delta|}{|h_k|} \ll 1, \qquad \text{for} \;\; k \in {\cal K}_i.
\end{align}
The proof is as follows: the condition $|\hat h_k| = |h_k + \Delta| \geq \sqrt{g_{\sf th}}$, for $k \in {\cal K}_i$ suggests that $|h_k| + |\Delta| \geq \sqrt{g_{\sf th}}$,  for $k \in {\cal K}_i$. Then we have $\frac{|h_k|}{|\Delta|} \geq \frac{\sqrt{g_{\sf th}}}{|\Delta|} - 1 \geq \frac{\sqrt{g_{\sf th}}}{|\Delta_{\max}|} - 1 \gg 1$ as desired.

\begin{figure*}[hh]
\begin{align} \label{app:eq:conditioned_err_prob_imperfect}
P^{\sf err}_i (K_i)  \leq
\l\{\begin{aligned} 
&  \frac{1}{\sqrt{K_i}S_i} +  \l(\frac{\sigma_z}{K_i\sqrt{\rho_0}} + \frac{2\sigma_\Delta}{\sqrt{K_i} \sqrt{\sqrt{g_{\sf th}} - \Delta_{\max}}}\r) \l( \frac{1}{S_i} + \frac{1}{2}\r), && K_i \geq 1, \\
&  \frac{1}{2}, && K_i = 0. 
\end{aligned}
\r.
\end{align}
\hrule
\vspace{-3mm}
\end{figure*}

From \eqref{app:del_over_hk}, by Taylor expansion we have $\frac{1}{1 + \frac{\Delta}{h_k}} = 1 - \frac{\Delta}{h_k} + O\l( \l(\frac{\Delta}{h_k}\r)^2 \r)$. By ignoring the high order term,  $\tilde g_i^{(n)}$ in \eqref{eq:aggregated_gradient_fading_im_CSI} can be approximated as 
\begin{align}\label{eq:aggregated_gradient_fading_im_CSI_2}
\tilde g_i^{(n)}  \approx \sum_{k \in {\cal K}_i}\sqrt{\rho_0}  \tilde g_{k,i}^{(n)} + z_i  - {I_i},
\end{align}
where ${I_i} = \sum_{k \in {\cal K}_i} \frac{\Delta}{h_k} \sqrt{\rho_0} \tilde g_{k,i}^{(n)}$ captures the error introduced by the imperfect CSI. Note that $\tilde g_{k,i}^{(n)}$ takes only the binary values of $+1$ and $-1$, 
and $|h_k| \geq \sqrt{g_{\sf th}} - \Delta_{\max}$, for $k \in {\cal K}_i$. Conditioned on $K_i$,  we can bound the variance of $I_i$ by
\begin{align}
{\sf Var}(I_i) \leq \frac{\rho_0 K_i \sigma_{\Delta}^2}{\sqrt{g_{\sf th}} - \Delta_{\max}}.
\end{align}

By noting the similarity between \eqref{eq:aggregated_gradient_fading_im_CSI_2} and \eqref{eq:aggregated_gradient_AWGN}, and taking ${I_i}$ as an additional noise introduced by the imperfect CSI, we can apply the same machinery presented in Appendix \ref{app:lemma:error_prob_AWGN} and the argument in the proof of Lemma \ref{lemma:conditioned_err_prob} to derive the conditional bit error probability for the imperfect CSI case as follows. The detailed proof is skipped due to space constraint. 

\begin{lemma}\label{lemma:conditioned_err_prob_imperfect} \emph{Consider the fading channel with truncated channel inversion using imperfect CSI. The bit error probability conditioned on the size of  the transmitting set $K_i$ is given by \eqref{app:eq:conditioned_err_prob_imperfect} as shown on the top of the current page.}

\end{lemma}

Then by the law of total probability and using the results in \eqref{eq:prob_of_K_i},
the unconditional error probability $P^{\sf err}_i = P^{\sf err}_i(K_i = 0)  \cdot \mathbb{P} (K_i = 0) + P^{\sf err}_i(K_i \geq 1) \cdot \mathbb{P} (K_i \geq 1)$ is  bounded by 
\begin{multline}\label{eq:unconditioned_error_prob_imperfect}
P^{\sf err}_i  \leq 
\frac{1}{2} (1-\alpha)^K +  \sum_{k=1}^K {K \choose k} \alpha^k (1-\alpha)^{K-k} \times \\
\l[ \frac{1}{\sqrt{k}S_i} + \l(\frac{\sigma_z}{k\sqrt{\rho_0}} + \frac{2\sigma_\Delta}{\sqrt{k} \sqrt{\sqrt{g_{\sf th}} - \Delta_{\max}}}\r) \l( \frac{1}{S_i} + \frac{1}{2}\r) \r].
\end{multline}
The above expression can be further simplified by applying Lemma \ref{lemma:two_inequalities} as presented below.
\begin{lemma}\label{lemma:unconditioned_err_prob_imperfect} \emph{Consider the fading channel with truncated channel inversion power control with imperfect CSI. The unconditional bit error probability is given by}
\begin{multline}
P^{\sf err}_i = \frac{1}{2} (1-\alpha)^K +    \frac{\sqrt{6}}{\sqrt{\alpha K}} \times \nn\\
\l[\frac{1}{S_i} + \l(\frac{2}{S_i} +1\r) \frac{\sigma_{\Delta}}{\sqrt{\sqrt{g_{\sf th}}-\Delta_{\max}}} \r] + \frac{2}{\alpha K} \frac{\sigma_z}{\sqrt{\rho_0}} \l( \frac{1}{S_i} + \frac{1}{2}\r) .
\end{multline}
\end{lemma}
With Lemma \ref{lemma:unconditioned_err_prob_imperfect} at hand, the desired result in Theorem \ref{theo:imperfect_CSI} can be easily derived by substituting Lemma \ref{lemma:unconditioned_err_prob_imperfect} into \eqref{eq:conditioned_expected_improvement} and following the same machinery presented in \eqref{eq:conditioned_expected_improvement_2}-\eqref{eq:bounded_improvement}.

\vspace{-3mm}
\bibliographystyle{ieeetr}
\bibliography{BibDesk_File_v2}
\vspace{-2mm}

\vspace{-30pt}
\begin{IEEEbiography}[{\includegraphics[width=1in,height=1.25in,clip,keepaspectratio]{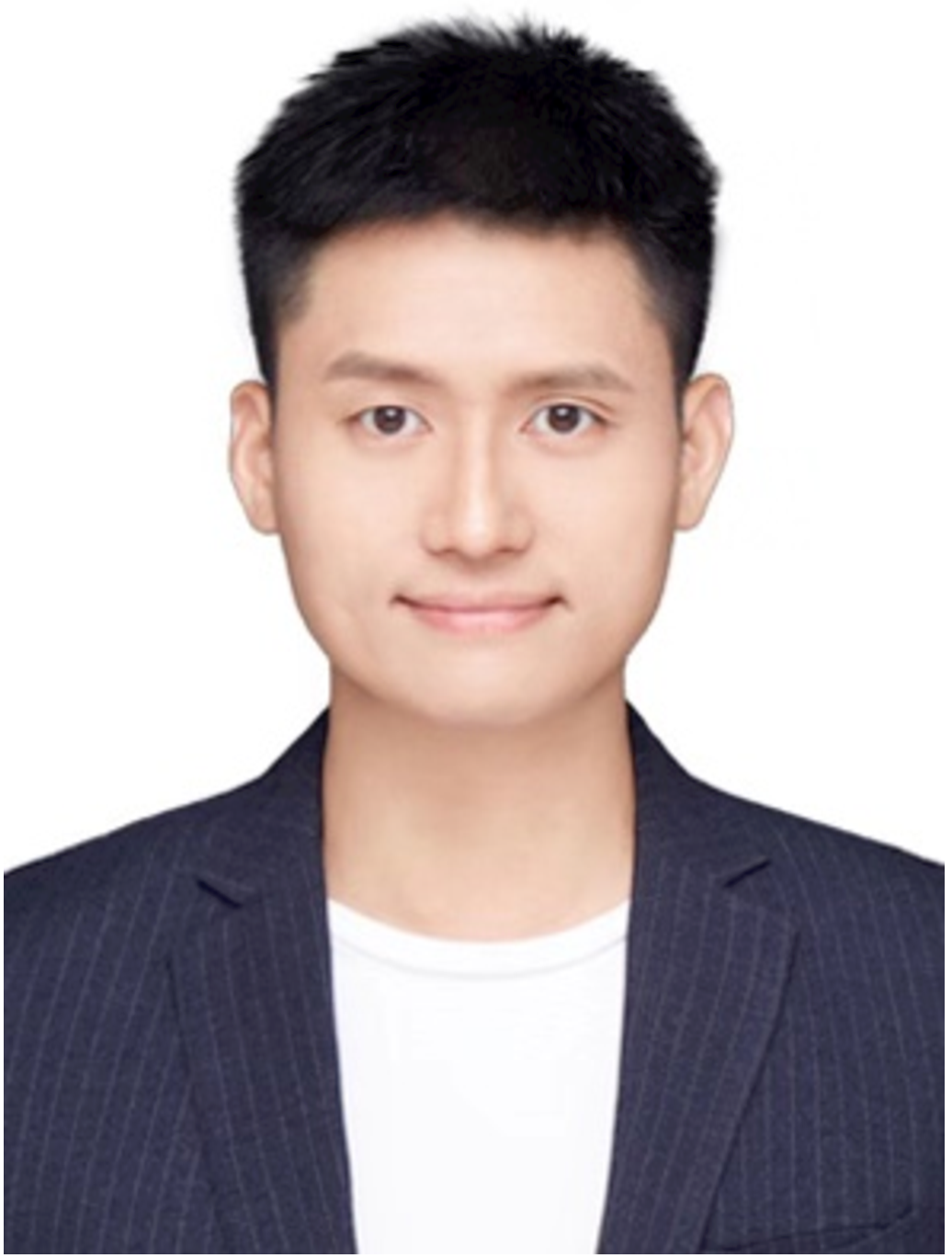}}]{Guangxu Zhu} received the B.S. and M.S. degrees from Zhejiang University and the Ph.D. degree from the University of Hong Kong, all in electronic and electrical engineering. He is now a research scientist with the Shenzhen Research Institute of Big Data. His research interests include edge intelligence, distributed machine learning, 5G technologies such as massive MIMO, mmWave communication, and wirelessly powered communication. He is a recipient of the Hong Kong Postgraduate Fellowship (HKPF), Outstanding Ph.D. Thesis Award from HKU, and a Best Paper Award from WCSP 2013. He was invited to be a co-chair for the "MAC and cross-layer design" track in IEEE PIMRC 2021.
\end{IEEEbiography}

\vspace{-30pt}
\begin{IEEEbiography}[{\includegraphics[width=1in,height=1.25in,clip,keepaspectratio]{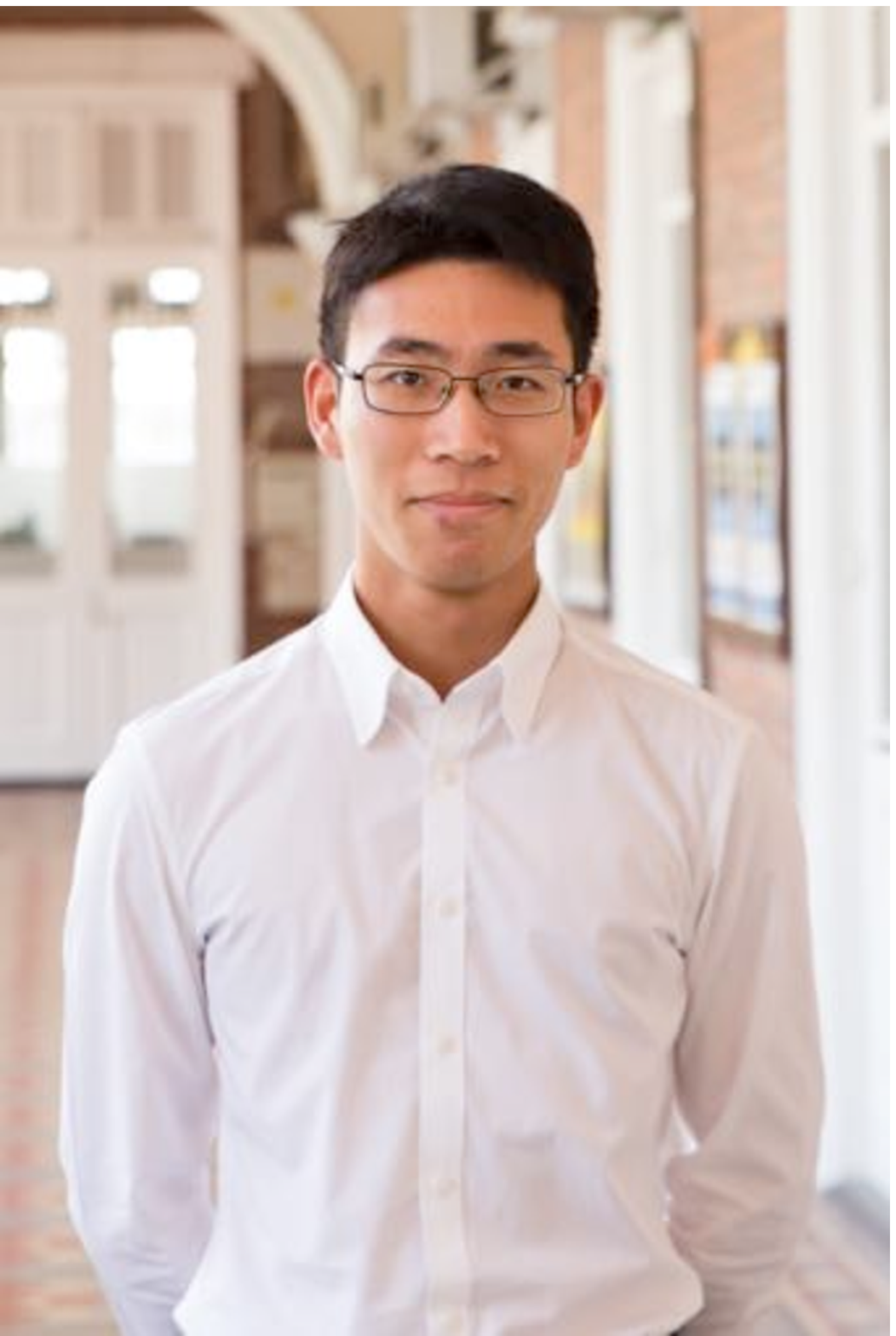}}]{Yuqing Du}
received the B.Eng. degree in Electrical Engineering from Harbin Engineering University, Harbin, China, in 2016, and the Ph.D. degree in Electronics and Electrical Engineering, The University of Hong Kong, Hong Kong, in 2020. His research interests include edge intelligence and distributed machine learning.
\end{IEEEbiography}

\vspace{-30pt}

\begin{IEEEbiography}[{\includegraphics[width=1in,height=1.25in,clip,keepaspectratio]{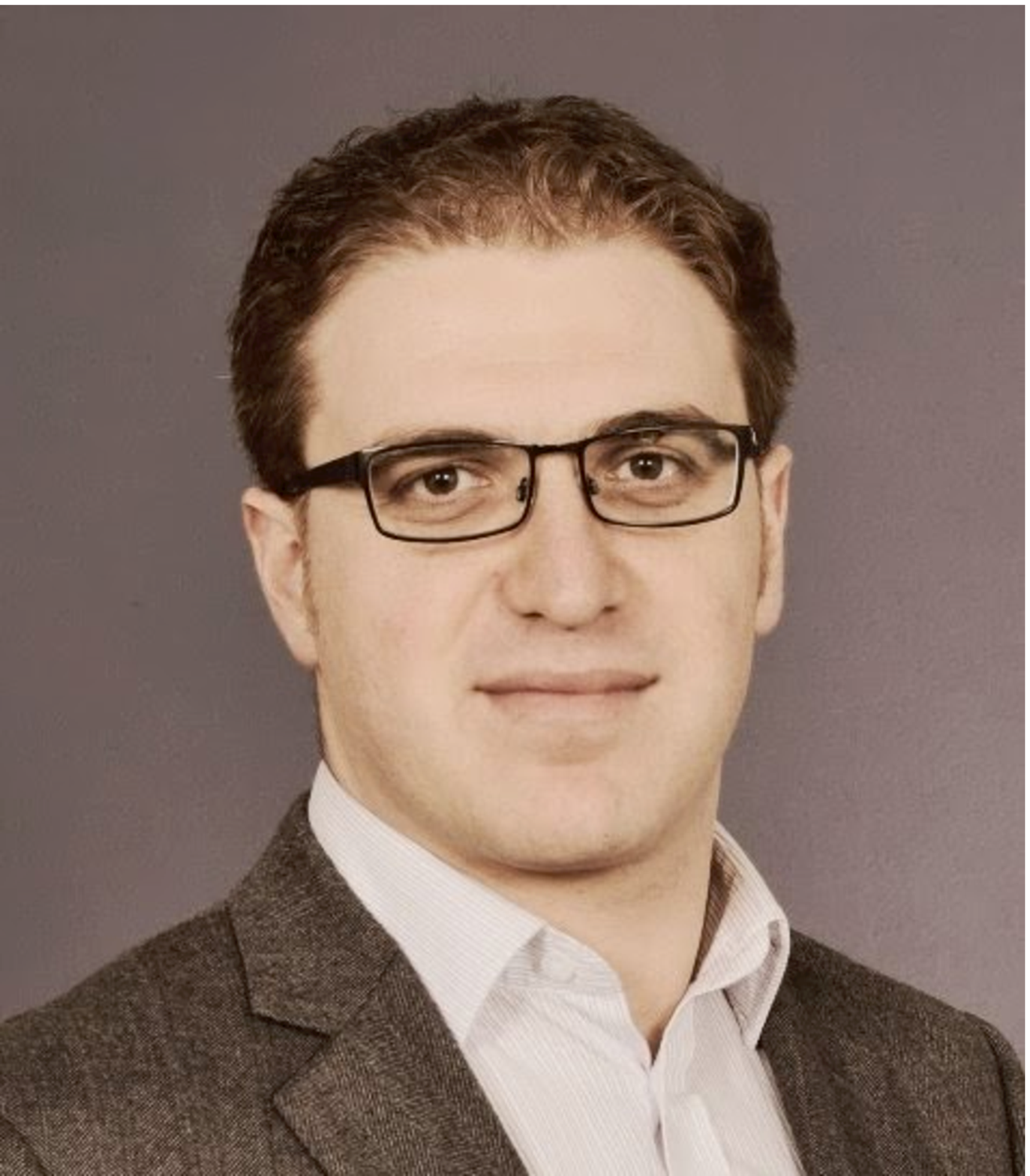}}]{Deniz G\"und\"uz} (Senior Member, IEEE) received M.S. and Ph.D. degrees in electrical engineering from NYU Tandon School of Engineering (formerly Polytechnic University) in 2004 and 2007, respectively. After his PhD, he served as a postdoctoral research associate at Princeton University, and as a consulting assistant professor at Stanford University. From Sep. 2009 until Sep. 2012 he served as a research associate at CTTC in Barcelona, Spain. ln Sep. 2012, he joined the Electrical and Electronic Engineering Department of Imperial College London, UK, where he is currently a Professor in Information Processing, serves as the deputy head of the Intelligent Systems and Networks Group, and leads the Information Processing and Communications Laboratory (IPC-Lab). He is also a part-time faculty member at the University of Modena and Reggio Emilia, and has held visiting positions at University of Padova (2018-2020) and Princeton University (2009-2012).

His research interests lie in the areas of communications and information theory, machine learning, and privacy. Dr. G\"und\"uz is an Area Editor for the IEEE Transactions on Communications and the IEEE Journal on Selected Areas in Communications (JSAC) - Special Series on Machine Learning in Communications and Networks. He also serves as an Editor of the IEEE Transactions on Wireless Communications. Previously, he served as an Editor for the IEEE Transactions on Green Communications and Networking (2016-2020), and the Transactions on Communications (2013-18). He is a Distinguished Lecturer for the IEEE Information Theory Society (2020-22). He is the recipient of the IEEE Communications Society - Communication Theory Technical Committee (CTTC) Early Achievement Award in 2017, a Starting Grant of the European Research Council (ERC) in 2016, and the IEEE Communications Society Best Young Researcher Award for the Europe, Middle East, and Africa Region in 2014. He coauthored papers that received best paper awards at 2019 IEEE GlobalSIP and 2016 IEEE WCNC, and Best Student Paper Awards at 2018 IEEE WCNC and 2007 IEEE ISIT. 
\end{IEEEbiography}

\vspace{-30pt}
\begin{IEEEbiography}[{\includegraphics[width=1in,height=1.25in,clip,keepaspectratio]{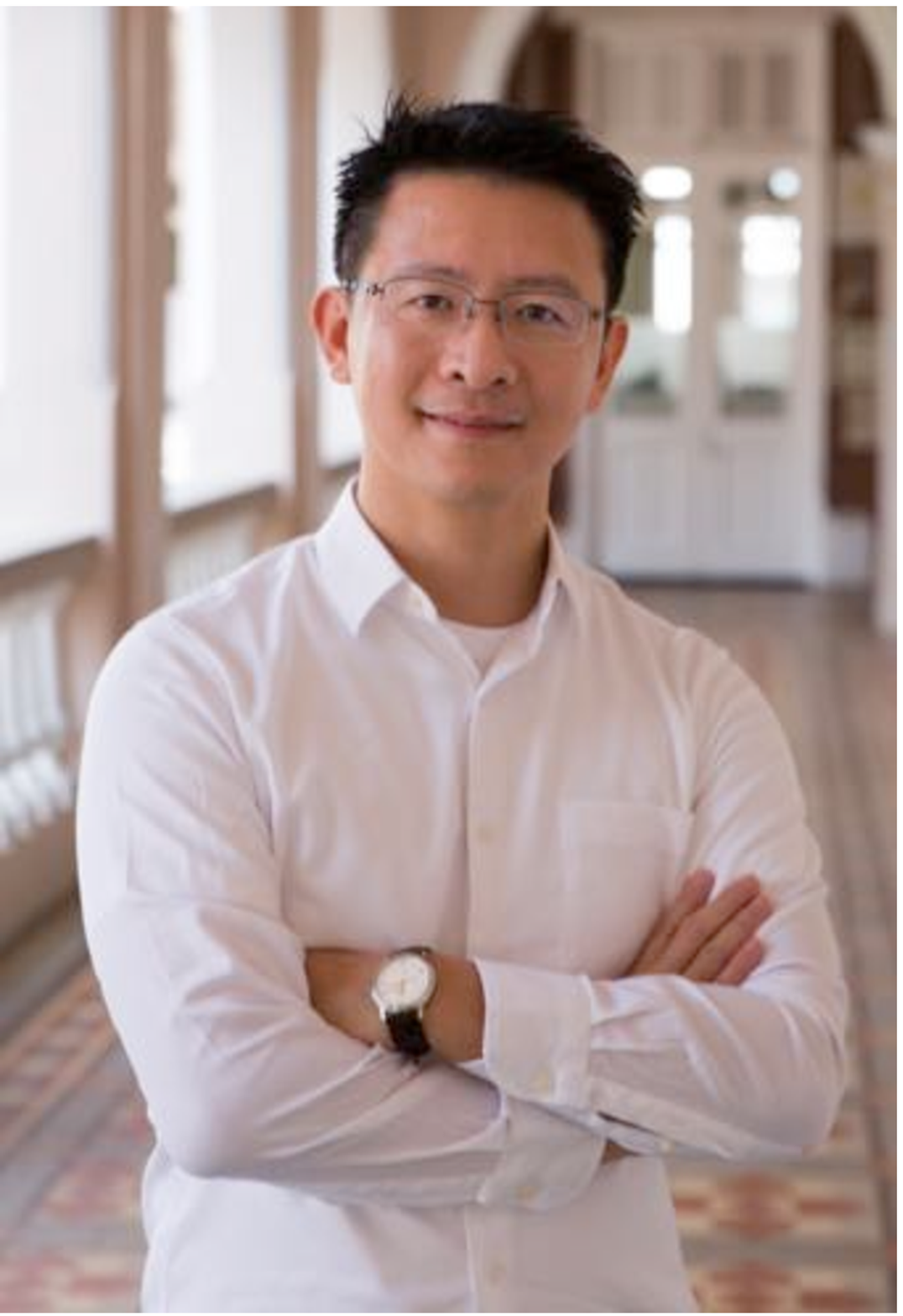}}]{Kaibin Huang} (Senior Member, IEEE) received the B.Eng. and M.Eng. degrees from the National University of Singapore, and the Ph.D. degree from The University of Texas at Austin, all in electrical engineering. Presently, he is an associate professor in the Dept. of Electrical and Electronic Engineering at The University of Hong Kong. He received the IEEE Communication Society’s 2019 Best Tutorial Paper Award, 2015 Asia Pacific Best Paper Award, and 2019 Asia Pacific Outstanding Paper Award as well as Best Paper Awards at IEEE GLOBECOM 2006 and IEEE/CIC ICCC 2018.  Moreover, he received an Outstanding Teaching Award from Yonsei University in S. Korea in 2011. He has served as the lead chairs for the Wireless Comm. Symp. of IEEE Globecom 2017 and the Comm. Theory Symp. of IEEE GLOBECOM 2014 and the TPC Co-chairs for IEEE PIMRC 2017 and IEEE CTW 2013. He is an Associate Editor for IEEE Transactions on Wireless Communications and Journal on Selected Areas in Communications (JSAC), and an Area Editor for IEEE Transactions on Green Communications and Networking. Previously, he has also served on the editorial board of IEEE Wireless Communications Letters. Moreover, he has guest edited special issues for IEEE JSAC,  IEEE Journal on Selected Topics on Signal Processing, and IEEE Communications Magazine. He is an IEEE Distinguished Lecturer and an ISI Highly Cited Researcher. 
\end{IEEEbiography}

\end{document}